\documentclass[english,prodmode,acmtosem]{acmsmall}
\usepackage[utf8]{inputenc}
\usepackage{babel}
\usepackage{fontenc}
\usepackage{graphicx}
\usepackage{booktabs}
\usepackage{subfig}
\usepackage{color}
\usepackage[normalem]{ulem}
\usepackage{threeparttable} 

\usepackage{floatrow}
\DeclareNewFloatType{example}{name=Example}
\DeclareCaptionLabelFormat{example-nn}{#1}
\title{A Taxonomy for Requirements Engineering and Software Test
Alignment}
\author{M. UNTERKALMSTEINER \affil{Blekinge Institute
of Technology} R. FELDT \affil{Chalmers University of
Technology and Blekinge Institute of Technology} T. GORSCHEK \affil{Blekinge
Institute of Technology}}
\begin{abstract}
Requirements Engineering and Software Testing are mature areas and have seen a 
lot of research. Nevertheless, their interactions have been sparsely explored 
beyond the concept of traceability. To fill this gap we propose a definition of 
requirements engineering and software test (REST) alignment, a taxonomy that 
characterizes the methods linking the respective areas, and a process to assess 
alignment. The taxonomy can support researchers to identify new opportunities 
for investigation, as well as practitioners to compare alignment methods and 
evaluate alignment, or lack thereof.
We constructed the REST taxonomy by analyzing alignment methods published in
literature, iteratively validating the emerging dimensions. The resulting 
concept of an information dyad characterizes the exchange of information 
required for any alignment to take place.
We demonstrate use of the taxonomy by applying it on five in-depth cases
and illustrate angles of analysis on a set of thirteen alignment methods. In 
addition we developed an assessment framework (REST-bench), applied it in an 
industrial assessment, and showed that it, with a low effort, can identify 
opportunities to improve REST alignment.
Although we expect that the taxonomy can be further refined, we believe that 
the information dyad is a valid and useful construct to understand alignment.
\end{abstract}
\acmformat{Unterkalmsteiner, M., Feldt, R., and Gorschek, T. 2013. A Taxonomy
for Requirements Engineering and Software Test Alignment.}
\begin{document}
\begin{bottomstuff}
Author's addresses: M. Unterkalmsteiner and T. Gorschek, Software Engineering
Research Lab, School of Computing, Blekinge Institute of Technology; R. Feldt,
Department of Computer Science and Engineering, Chalmers University of
Technology.
\end{bottomstuff}
\maketitle

\section{Introduction}\label{sec:Introduction}
Industrial-scale software development is an undertaking
that requires judicious planning and coordination of the involved resources. The
inception, design, implementation, examination and maintenance of a software
product~\cite{scacchi_process_2001} are a team effort, organized and executed to
satisfy the product customer. Following the separation of concerns principle,
software life-cycle models distinguish between different phases or
activities in the production of software, linking them by feed-forward and
feed-back loops~\cite{madhavji_process_1991}. This separation reduces the
complexity of each single phase or activity, however at the same time poses
needs for an efficient and effective coordination. 

In this paper, we investigate two phases in the software development
life-cycle, requirements engineering (RE) and software testing (ST), that
benefit particularly from a coordinated
functioning~\cite{graham_requirements_2002}. Several prominent researchers have
called for more research towards this goal. At FoSE 2007,
\citeN{cheng_research_2007} called for a stronger collaboration
between RE and researchers and practitioners from other software engineering
fields to improve requirements knowledge and downstream development.
\citeN{bertolino_software_2007} summarized current challenges and goals
in software testing research, pointing out the rising importance of a more
holistic approach to ST which takes advantage of the overlaps between different
research disciplines.
Recent research shows that the study of the synergies between RE and ST
are important and of particular interest for
industry~\cite{uusitalo_linking_2008,post_linking_2009,%
sabaliauskaite_challenges_2010}.

Despite these advancements and its relevance for practitioners, there is
still a lack of research that aims at understanding, characterizing and
communicating methods that align requirements engineering and software test. By
studying methods for RE and ST alignment we intend to fill this gap. This paper
does not aim at providing a systematic and exhaustive state-of-the-art survey of
RE or ST research, but rather forms the foundation, through a taxonomy, to 
classify and characterize alignment research and solutions that focus on the 
boundary between RE and ST. The REST taxonomy also functions as an engine for 
REST-bench, an alignment assessment framework.

With \textit{alignment} we mean the \textit{adjustment of RE and ST efforts for
coordinated functioning and optimized product development.}
Depending on the context, alignment can be understood as an activity
or as a state. Alignment-as-activity pertains to the act of \textit{adjusting
or arranging} efforts involved in RE and ST so that they work better together. 
To improve our understanding of such activities, we developed the REST 
taxonomy. Alignment-as-state, on the other hand, refers to the condition of
RE and ST efforts having \textit{established} a coordinated functioning. In 
order to evaluate the state of alignment we developed REST-bench which acts as 
an assessment framework and is based on the REST taxonomy.
Independently from the context, the above definitions imply that a higher
degree of alignment enables higher effectiveness and efficiency in
product development and/or maintenance.

In this paper we study RE and ST alignment with the purpose of
\begin{longitem}
 \item \emph{Characterization} of RE and ST alignment methods, providing
researchers and practitioners a common vocabulary
 \item \emph{Analysis} of RE and ST alignment methods, providing researchers
means to preemptively identify weaknesses and suggest improvements
 \item \emph{Industrial assessment} of RE and ST alignment, providing
practitioners a lightweight framework (REST-bench, powered by the REST 
taxonomy) to identify
misalignment
\end{longitem}

The remainder of the paper is structured as follows. In
Section~\ref{sec:Background} we discuss the relationship between requirements
engineering and software testing in more detail and illustrate related work. In
Section~\ref{sec:TheTaxonomy} we present the REST taxonomy, accompanied with an
example of its application, and classify thirteen alignment methods. In 
Section~\ref{sec:Validation} we illustrate the process followed for 
constructing and validating the taxonomy. In 
Section~\ref{sec:AlignmentAsActivity} we analyze the classified methods by the 
means the REST taxonomy provides. We introduce REST-bench, which we applied in 
an industrial case study at Ericsson AB, in Section~\ref{sec:AlignmentAsState}. 
The paper concludes with Section~\ref{sec:Conclusion}, pointing out directions 
for future work.

\section{Background and Related work}\label{sec:Background}
\subsection{The need for alignment}
Software development consists of transitions from system concept, requirements
specification, analysis and design, implementation, and test and
maintenance~\cite{laplante_what_2007}. This abstraction holds for both
plan driven process models (e.g. spiral~\cite{boehm_spiral_1988} and
evolutionary~\cite{naumann_prototyping:_1982}, and the unified process
model~\cite{kruchten_rational_2000}), as well as and Agile models, although to a
lesser extent in the latter category as activities may be blended, eliminating
transitions altogether (e.g. in eXtreme Programming~\cite{beck_embracing_1999}).

Looking at the V-Model, which originates from system
engineering~\cite{forsberg_relationship_1991,brohl_v-_1995} and was adopted
in software engineering~\cite{pfleeger_software_2009}, high-level testing is
often depicted as the Verification~\&~Validation activity to requirements
elicitation, analysis and specification. As such, this connection between
requirements engineering and testing is a key part of our software engineering
knowledge. Still, this connection is not considered in detail as a collective
concept in our research activities. On the other hand, an abundance of software
technologies, models and frameworks have been developed to ease the transition
of software development phases, to bridge the gap between them, and to
\emph{align} the intentions and activities therein, for example, between
requirements and software architecture/design
(\cite{kop_conceptual_1998,amyot_bridging_2001,hall_relating_2002}), software
architecture/design and implementation
(\cite{murphy_software_2001,elrad_aspect-oriented_2002,aldrich_archjava:_2002}),
and software architecture/design and testing
(\cite{muccini_using_2004,samuel_automatic_2007}).

However, aligning requirements engineering and software testing is a less
explored territory, although it would be beneficial to recognize the inherent
link between them~\cite{graham_requirements_2002}. The need for RE and ST
alignment is emphasized by the difficulty to design, implement and
maintain large software systems. The increase in complexity of the problem
space, i.e. requirements, increases also the complexity of the software solution
~\cite{glass_sorting_2002}, making therefore the testing more involved. Benefits
of a strengthened  link between RE and ST are, for example, improved product
quality~\cite{uusitalo_linking_2008}, cost-effective
testing~\cite{miller_case_2010,flammini_automatic_2009}, high quality
test-cases~\cite{de_santiago_junior_generating_2012}, and early discovery of
incomplete requirements~\cite{siegl_model_2010}. 

The means by which RE and ST alignment can be achieved, include (but are not
limited to) methods or processes that establish and maintain requirements
to test traceability links~\cite{gotel_analysis_1994,ramesh_toward_2001}, use
requirements as a driver to develop tests (e.g. by formulating testable
contracts~\cite{melnik_executable_2006,martin_tests_2008}, use model-based
testing~\cite{utting_taxonomy_2011}), or organize development teams in an
effective manner (e.g. by forming cross-functional
teams~\cite{marczak_how_2011}).

The means of achieving alignment are diverse in terms of the assumptions they
make, their prerequisites on the organizational environment, and the investments
they require. Effectively searching, selecting and applying instruments to
improve RE and ST alignment is therefore a challenge for practitioners but also
for researchers in advancing the state-of-the-art. \citeN{uusitalo_linking_2008}
conducted interviews at five Finnish software organizations and elicited
practices, such as tester participation in requirements reviews, and
requirements to test traceability, that aim to bridge the gap between RE and ST.
\citeN{post_linking_2009} explored how the impact of requirements changes, and
the subsequent effort in adapting test cases, can be reduced by scenario-based
requirements formalizations. In an interview study with software practitioners
occupying roles as quality control leaders, requirements process managers and
test leaders, Sabaliauskaite et al. identified several obstacles in aligning
requirements engineering and testing. Barriers exist in the organizational
structure, processes and cooperation between people, and are aggravated by tool
deficiencies and challenges in change
management~\cite{sabaliauskaite_challenges_2010}.

The connections between RE and ST are both clear and numerous, and the 
potential benefits in increasing the coordination between them are large. 
Therefore it is essential that we increase our understanding through the study 
of these connections, and treat them as a collective and not as individual, 
isolated areas and approaches. Our main aim in this paper is to systematically 
create a basis for such an understanding. In order to characterize the 
phenomenon of alignment between RE and ST we developed therefore the REST 
taxonomy. 

\subsection{Alignment vs. Traceability}\label{sec:Traceability}
The concept of traceability, which exists since the dawn of the software 
engineering discipline~\cite{randell_towards_1968}, is not associated with a 
particular goal, but is a quality attribute of the artifacts produced in 
software development. The IEEE Standard Glossary of Software Engineering 
Terminology~\cite{ieee_ieee_1990} defines traceability as ``the degree to which 
a relationship can be established between two or more products of the 
development process, especially products having a predecessor-successor or 
master-subordinate relationship to one another [...]''. Gotel and Finkelstein 
provide a similar definition of requirements traceability as ``the ability to 
describe and follow the life of a requirement''~\cite{gotel_analysis_1994}, 
which complies with the notion of traceability being a work product quality 
attribute. 

Research into traceability indicates that good traceability supports impact 
analysis~\cite{gotel_analysis_1994,ramesh_toward_2001,damian_requirements_2005,%
uusitalo_linking_2008} and lowers test and maintenance 
costs~\cite{watkins_why_1994,kukkanen_applying_2009}. On the other hand, high 
quality traces are expensive to establish and 
maintain~\cite{cleland-huang_event-based_2003}, leading to the investigation of 
means to automate the trace recovery 
process~\cite{de_lucia_recovering_2007,hayes_requirements_2007}. 

We defined alignment as a goal-directed concept, i.e. the 
adjustment of RE and ST efforts for coordinated functioning and 
optimized product development. As such, high quality traces may 
contribute to an improved alignment, are however not the only solution 
candidates achieving our goal of alignment. Thus, traceability can be a method 
to achieve alignment, but the REST taxonomy focuses on the alignment phenomena 
itself and how methods for alignment (which might build on traceability) can be 
classified.

\subsection{The purpose of taxonomies}\label{sec:TaxonomyTypes}
Creating taxonomies of objects or concepts has been a basic scientific tool
since early work by the Swedish botanist Carl von
Linné~\cite{linnaei_systema_1735}. Taxonomies are means to structure, advance
the understanding, and to communicate
knowledge~\cite{glass_contemporary_1995,kwasnik_role_1999}. When the
understanding in a certain area advances, concepts and relationships between
them emerge that allow for a structured representation of these concepts. Being
able to communicate that knowledge provides the opportunity to further advance
research~\cite{kwasnik_role_1999}. 
Kwasnik also points out the importance of taxonomies as theory developing tools.
Classification schemes enable the display of theory in an useful way and serve,
similar to theories, as drivers for inquiry~\cite{kwasnik_role_1992}. Thus,
the development of taxonomies is essential to document theories which
accumulate knowledge on Software Engineering
phenomena~\cite{sjoberg_future_2007}.

\subsection{Taxonomies in Software Engineering}\label{sec:TaxonomiesSE}
The Guide to the Software Engineering Body of Knowledge (SWEBOK) is an attempt
to characterize the software engineering discipline and to provide a structured
access to its body of knowledge~\cite{bourque_guide_2004}. As such, SWEBOK can
be seen as a taxonomy that covers knowledge areas relevant to software
engineering, promoting the structured communication of this discipline.
Similarly, \citeN{glass_research_2002} provide a taxonomy on the
research in software engineering, although its main purpose is to structure and
to position past research. Blum's taxonomy of software development
methods~\cite{blum_taxonomy_1994} is more narrow in scope and, similar
to Glass et al., aims at structuring rather than communicating the
knowledge on software development methods. 

Further examples of specialized taxonomies, i.e. with a narrow scope, are
\citeN{buckley_towards_2005} on mechanisms of software change, 
\citeN{svahnberg_taxonomy_2005} on variability realization techniques, and
\citeN{mehta_towards_2000} on software component connectors.

\subsection{Developing Taxonomies}\label{sec:DevelopingTaxonomies}
The development of a taxonomy can be approached in two different ways, top-down
and bottom-up~\cite{glass_research_2002}. In the top-down or
enumerative~\cite{broughton_essential_2004} approach, the classification scheme
is defined a-priori, i.e. a specific structure and categories are established
that aim to fulfill the purpose of the taxonomy. The created classification
scheme is thereby often a composition of previously established schemata
(e.g.~\cite{glass_research_2002,avizienis_basic_2004,bunse_taxonomy_2006}), or
the result of the conceptual analysis of a certain area of interest
(e.g.~\cite{svahnberg_taxonomy_2005,utting_taxonomy_2011}). The strength of
this approach is that the taxonomy is built upon existing knowledge structures,
allowing the reuse of established definitions and categorizations and hence
increasing the probability of achieving an objective classification procedure.

On the other hand, the bottom-up or
analytico-synthetic~\cite{broughton_essential_2004} approach is driven by the
sampling of subjects from the population of interest and the extraction of
patterns that are refined into a classification scheme. For example,
\citeN{vegas_maturing_2009} extended existing unit-testing classifications by
systematically studying the Software Engineering literature, supplemented by
gathering the expert judgment of researchers and practitioners in the testing
area. The strength of this approach is that new, not yet classified,
characteristics may emerge and enrich existing taxonomies.

The goal of the taxonomy presented in this paper is to classify methods
that bridge the gap between requirements engineering and software testing
activities. There exists a rich knowledge base for both RE and ST, and
taxonomies for classifying aspects in each area already exist. Following a
top-down approach and amalgamating concepts, definitions and categorizations
from these separate areas into a taxonomy of RE and ST alignment seemed to us
unlikely to succeed. Even though the respective areas are mature and have seen a
lot of research, their interplay and connections have been less explored. Hence
we chose to construct the taxonomy in a bottom-up fashion, validating the
emerging classification scheme throughout the process (see
Section~\ref{sec:Validation}).

\section{The REST Taxonomy}\label{sec:TheTaxonomy}
\begin{figure}
 \subfloat[Information dayd]{\label{fig:InformationDyad}
  \includegraphics[scale=1.0]{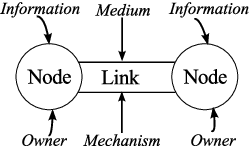}
  }
  \subfloat[Dyad structure of a method]{\label{fig:ExampleDyadStructure}
   \includegraphics[scale=1.0]{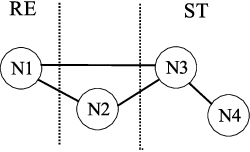}
  }
  \subfloat[Method classification]{\label{fig:ExampleClassification}
   \includegraphics[scale=1.0]{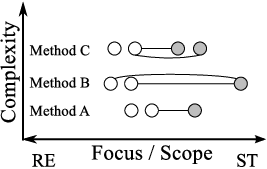}
  }
 \caption{Anatomy of the REST taxonomy}\label{fig:RESTAnatomy}
\end{figure}

When developing a taxonomy one has to consider its
\textit{purpose}~\cite{glass_contemporary_1995}. A specific taxonomy is designed
to accommodate a single, well-defined purpose. On the other hand, the
structure of general taxonomy is not imposed by a specific
purpose~\cite{glass_contemporary_1995} and is hence applicable in various
circumstances. As we defined earlier in Section~\ref{sec:Introduction},
alignment can be understood as an activity or a state. We therefore designed the
structure of our taxonomy to accommodate both aspects of the alignment
definition. From the alignment-as-activity perspective, the REST taxonomy 
can be used to analyze and categorize alignment methods described in 
literature. From the alignment-as-state perspective, the REST taxonomy 
serves as an analysis aid in project and process assessment. The method we 
developed for process assessment (which connects to alignment-as-state), 
REST-bench, is described and illustrated through a case study in 
Section~\ref{sec:AlignmentAsState}.

Figure~\ref{fig:RESTAnatomy} provides an overview of the REST taxonomy. The 
taxonomy is centered around our observation (see 
Section~\ref{sec:TaxonomyConstruction}) that the alignment of RE and ST implies 
some sort of information linkage or transfer between two entities involved in 
either process area. In essence, if there is no exchange of information, at 
least at some point in time, no alignment can take place or be achieved. Thus, 
characterizing such exchanges are key in a general taxonomy. In order to 
describe this phenomenon we devised the concept of an information dyad, 
representing the central unit of analysis in the taxonomy.
The information dyad contains the \textit{criteria for differentiation and
description}~\cite{glass_contemporary_1995} used for the classification, the
second essential aspect in taxonomy development.
Figure~\ref{fig:InformationDyad} illustrates the components of an
information dyad. A \emph{node} is characterized by the type of information it
represents and an owner of that information. Two nodes are connected by a
\emph{link}, characterized by the type of mechanism establishing the link
between nodes, and the medium through which the link is realized.

The third important property of a taxonomy, besides its purpose and criteria for
differentiation, is the \textit{method of
classification}~\cite{glass_contemporary_1995}. The method should illustrate
and explain how objects under study are classified in a repeatable and
unambiguous manner. To this end we developed a process, summarized in
Table~\ref{tab:RESTProcess}, in which each step answers a
specific question. 

The objects under study are methods that may improve the alignment between RE 
and ST, published at conferences or in journals.
Hence, Step 1 in the process serves as a gatekeeper, asserting that the taxonomy
is applied on studies that can answer the questions asked in the following
steps. Steps 2.1, 2.2 and 2.3 aim at identifying the information dyads 
of the studied method, and characterizing the dyads by their components 
(information, medium and mechanism). The context in which the alignment method 
has been developed or applied is captured in Step 3.

Since an alignment method consists of one or more dyads, these dyads form a 
structure which characterizes the method 
(Figure~\ref{fig:ExampleDyadStructure}). In Step 4 we analyze the properties of 
the dyad structure which allows us in Step 5 to classify the methods according 
to their complexity and scope/focus (Figure~\ref{fig:ExampleClassification}).

The remainder of this section describes the process shown in 
Table~\ref{tab:RESTProcess}. To complement the description, we illustrate the
application of the taxonomy by self-contained examples, based on
\citeN{miller_case_2010} case study on their framework and method of
model-based testing of specifications and implementations. Note that section and 
figure numbers in the examples refer to~\citeN{miller_case_2010}.
Finally, we apply the taxonomy on 13 alignment methods in 
Section~\ref{sec:ProcessMethodClassification}.

\begin{table}[t]
\tbl{REST classification process\label{tab:RESTProcess}}{%
\begin{tabular}{lll}\toprule
Step & Question answered & Section\\ \midrule
1 & Does the study shed light on both RE and ST aspects? &
\ref{sec:ProcessRelevance}~Relevance\\
2 & What are the components of the information dyad? & 
\ref{sec:ProcessInformationDyad}~The information dyad\\
~2.1 & ~What type of information exists/is used in RE and ST? &
~\ref{sec:ProcessInformation}~Information\\ 
~2.2 & ~What type of medium connects the information? &
~\ref{sec:ProcessMedium}~Medium\\ 
~2.3 & ~What type of mechanism establishes the connection? &
~\ref{sec:ProcessMechanism}~Mechanism\\
3 & In which environment is the method situated? &
\ref{sec:ProcessContext}~Method context\\ 
4 & What is the structure of the identified dyads? &
\ref{sec:ProcessStructure}~Dyad structure properties\\ 
5 & How can the method be classified? &
\ref{sec:ProcessMethodClassification}~Method classification\\
\bottomrule
\end{tabular}}
\end{table}

\subsection{Relevance}\label{sec:ProcessRelevance}
The analyst (the person who applies the taxonomy) needs to decide whether the
study, and the described alignment method therein, qualifies to be classified
with the taxonomy. He bases his decision on three independent criteria:
\begin{longitem}
 \item \emph{Scope}: Since the taxonomy aims at characterizing links between
requirements engineering and software testing activities, the candidate study
should consider both areas in the discussion of the presented method. If, for
example, the focus of the study is on formal reviews of requirement
specifications, considers however also the effects of reviews on downstream
development including testing or discusses the involvement of quality assurance
personnel in reviews, the study is likely to be adequate for taxonomy
application. On the other hand, a comparison of different review techniques,
focused on identifying respective strengths and weaknesses alone, is likely not
to be adequate.
 \item \emph{Comprehensiveness}: A detailed report of the conducted study
reduces the analyst's leeway for interpretation when answering the questions
posed in Table \ref{tab:RESTProcess}. It is impossible to judge
the comprehensiveness of a publication a-priori, i.e. before reading it, but
since space restrictions of journals are less rigid than for conference or
workshop publications, they tend to exhibit more details on the conducted study.
 \item \emph{Rigor}: In case the publication includes a method evaluation,
rigor of reporting context, design and validity threats
\cite{ivarsson_method_2010} should be considered. A strong description of these
aspects supports the analyst in performing context identification (Step 3,
Section~\ref{sec:ProcessContext}).
\end{longitem}

The example for Step 1 shortly introduces the publication on which all
following examples in this section are based upon, and illustrates the
application of the above discussed criteria to assess relevance.
\begin{example}[t]
 \caption{Step 1 - Does the study shed light on both RE and ST
aspects?}\label{exa:Example-1}
\citeN{miller_case_2010} present and evaluate a framework
that aims at taking advantage of synergy effects between specification testing
and implementation testing. They argue that the effort spent in verifying the
formal specification of a software system can also contribute to the
verification of the implementation of that system. To this end, they introduce
testgraphs as a mean to model parts of the specification that require testing,
using them for the derivation of test sequences in both specification and
implementation testing.

The table below illustrates our assessment with respect to the relevance 
criteria we defined.

\begin{center}
\begin{tabular}{lp{5cm}p{5cm}}\toprule
Criterion & Strength & Weakness \\ \midrule
Scope & + includes both RE and ST activities (Section 3) & - derivation of 
formal specifications from requirements is not part of the framework (Figure 
2)\\
 & + consistency and conformance check between testgraph and formal 
specifications (Section 3.1) & \\
Comprehensiveness & + activities respectively roles are described (Section 3, 
Section 5) & \\
Rigor & + context (Sections 4.2-4.4) and design (Section 4.1, Section 7) 
described & - threats to the validity of the evaluation not discussed \\
 & + risks in the application of the framework are considered (Section 3.3) & \\
\bottomrule
\end{tabular}
\end{center}
Based on this assessment, we conclude that we can apply the REST taxonomy on 
the described method.
\end{example}

\subsection{The information dyad}\label{sec:ProcessInformationDyad}
The goal of this step is to identify the nodes and the link 
that characterize an information dyad (see Figure~\ref{fig:InformationDyad}). 
Note however that an alignment method can be described by more than one dyad, 
depending on the number of identified nodes. Hence we discuss dyad structures 
and their properties in Section~\ref{sec:ProcessStructure}. 

\subsubsection{Information}\label{sec:ProcessInformation}
An information dyad consists of two nodes and a connecting link. A node
describes an entity that has to be aligned, synchronized, brought into
agreement, with another entity. The nodes represent the different, primary
objects of information, while the link represents the fact that they are or
should affect one or both of each other.
To differentiate between nodes, we assign each node a name, characterizing its
purpose. We deliberately do not limit the definition of a node to the notion of,
for example, a phase in the software development life-cycle. A node could also
be an activity, e.g. formal inspections \textit{during} requirements analysis.
Although this allows for more flexibility, it also reduces the repeatability in
the classification of the alignment method.

A node is characterized by the information it contains and an owner who is the 
source of that information. In this work, we informally define information as a
\textit{coherent collection of related data that is created during software
development, often with a specific purpose in mind}. Later on in
Section~\ref{sec:ProcessMechanism} we will further refine the notion of the 
information concept, but for this step in the classification process this
operational definition is sufficient.

Information, according to the above definition, is created, recorded and used at
any point in time during software development, enabling product inception,
specification, implementation, verification and validation, and maintenance.
Typically it refers to development artifacts but it can also represent more
intangible but essential and purposely related knowledge of a developers or
testers~\cite{feldt_biomimetic_2002}, e.g. informal requirements as in the
Miller and and Strooper case (see the example for Step 2.1).
With this taxonomy we aim to capture in particular information that is shared
and aligned in RE and ST activities. This does however not restrict the
information content to RE or ST topics, e.g. technical requirements, feature
descriptions, priorities, test plans, strategies, scenarios, etc. Information
valuable for alignment can emerge from any phase in software development and
connect RE and ST activities. Hence, the task of the analyst is to carefully
study the described method and collect evidence for the existence of a node and
its characterizing information. Such evidence can be found in statements on used
or created artifacts or in descriptions of things that have been discussed or
communicated. The owner, the second attribute of a node, is responsible for
creating and/or maintaining the information. Depending on the organization of
the development process, the owner may be formally assigned to this
responsibility (i.e. by occupying a specific role or function) or, in case of
agile processes, depend on the employee's current activities.
\begin{example}[t]
 \caption{Step 2.1 - What type of information exists/is used in RE
and ST?}\label{exa:Example-2}
To answer this question, we focus first on identifying actors and the
information on which is acted upon. The framework description and the
illustration of the performed tasks in the GSM case study are thereby of
interest (Section 3, 3.1, 3.2, 5.1, 6.1 and 6.2). Remember that our ultimate 
goal in this step is to identify potential nodes that form one or more 
information dyads. Hence the other components, medium and mechanism, play a 
secondary role at this moment. Defining the characteristics of links too early 
in the process may inhibit the discovery of all relevant nodes. On the other 
hand, since Steps 2.1-2.3 are performed iteratively, refinements are still 
possible at a later moment.
The table below lists the identified nodes. In the following we will motivate
them and illustrate in which relation they stand to each other, i.e. define the
information dyads in this example.
\begin{center}
\begin{tabular}{llll}\toprule
ID & Node name & Information & Owner\\ \midrule
N1 & Requirements specification & Informal requirements & Req. engineer\\ 
N2 & Req. analysis (Implementation) & Formal specification & Req.
engineer\\ 
N3 & Req. analysis (Test) & Testgraphs & Tester \\
N4 & Specification test & Test sequence & Tester \\ 
N5 & Specification mapping & Spec. to impl. mapping & Tester \\
N6 & Testgraph mapping & Testgraph to impl. mapping & Tester \\
N7 & Implementation test & Test sequence & Tester \\
\midrule
\multicolumn{4}{l}{Dyads (6): N1-N2, N1-N3, N3-N4, N2-N5, N3-N6, N6-N7}\\
\bottomrule
\end{tabular}
\end{center}
The first node, N1, contains the fundamental information, i.e. informal
requirements, from which further artifacts are derived. The formal
specification is developed by the requirements engineer to aid design and
implementation. Hence we define requirements analysis (Implementation) as the
second node (N2). Similarly, the tester develops a testgraph, with associated
test cases, to aid the verification of the formal specification and the
implementation. Requirements analysis (Test) is therefore the third node (N3).
The dyads, N1-N2 and N1-N3, follow from the refinement performed by requirements
engineers and testers. The specification is tested by generating test sequences
(N4) from the testgraph, leading to dyad N3-N4.

Identifying the next nodes is rather challenging. Figure 2 identifies the test 
oracle and the implementation as further artifacts relevant for the framework. 
From the alignment perspective however, the interesting part of the framework 
is the mapping between specification respectively testgraph to the 
implementation, described in Sections 6.1 and 6.2. The reason why it is
interesting is that the tester needs to process and understand artifacts
developed by requirements engineers (formal specification) and developers
(implementation). Hence, specification mapping (N5) and testgraph mapping (N6)
are nodes of interest, leading to dyads N2-N5 and N3-N6 representing the
relationships of the mapping. The implementation test (N7), also based on the
derivation of a test sequence, is driven by the testgraph mapping, leading to
dyad N6-N7.
\end{example}

\subsubsection{Medium}\label{sec:ProcessMedium}
The Oxford English Dictionary defines the term medium as ``an intermediate
agency, instrument, or channel; a means; especially a means or channel of
communication or
expression''~\cite{oxford_english_dictionary_medium_2011}. The medium in
an information dyad describes how the information between two nodes is linked
together. This can be through a carrier of information, e.g. an artifact, or a
facilitator that enables the information transfer, e.g. a process. During the
development of the taxonomy we have identified a set of different media types:
\begin{itemize}
 \item Structured artifacts (e.g. documents, email, diagrams, database records);
they are usually persistent and searchable/indexed.
 \item Unstructured artifacts (audio, video); they are usually not
searchable/indexed.
 \item Tools that act as means to share, transfer or transform information
(e.g. modeling tools, language analysis tools).
 \item Process (one or more activities, can be performed repeatedly).
 \item Organization of work environment (co-location, role/responsibility
rotation).
\end{itemize}
The analyst can choose one of these media types if appropriate or introduce a
new type as the above set was derived only from a sample of alignment methods
studied and hence be incomplete.
\begin{example}[t]
 \caption{Step 2.2 - What type of medium connects the
information?}\label{exa:Example-3}
For each dyad identified in Step 2.1 we now define their linking medium. 

In the dyad Requirements specification - Requirements analysis
(Implementation), a requirements engineer is responsible for deriving the
formal specification from informal requirements. We have to assume that this
derivation is performed manually, following a certain process, since the
framework description does not explain this step in detail. Hence, for the dyad
N1-N2 we declare the medium to be a process.

Similarly, testgraphs are derived by a tester from the informal requirements,
represented by the dyad Requirements specification - Requirements analysis
(Test). Also here, the derivation is a series of activities (define testgraph
and associated test cases, measure specification coverage) that follows standard
testing heuristics. Hence we declare also in the dyad N1-N3 the medium to be a 
process. The generation of test sequences (N4) is supported by a tool for the 
editing the graph and executing tests, leading to the conclusion that the 
medium in dyad N3-N4 is a tool.

Both in dyad N2-N5 and N3-N6, in which mappings between a model (specification
respectively testgraphs) and the implementation are created, the link medium is
a process. The tester implements wrapper classes for the classes under test
(dyad N2-N5), linking state, operations, input and output, and return values
from the implementation to the corresponding entities in the specification. In
dyad N3-N6, the tester performs a similar task by implementing a driver class
that calls for each traversed node and arc in the testgraph the appropriate
operation in the wrapper class. Although both mappings can be potentially
created automatically, such a tool is currently not available in the framework. 
On the other hand, the generation of test sequences for the implementation test 
(N7), is tool supported. Hence the medium in dyad N6-N7 is a tool.
\end{example}

\subsubsection{Mechanism}\label{sec:ProcessMechanism}
The mechanism component of a dyad link characterizes the way in which 
information is carried, eventually changing its purpose, from one node to the 
next. We assume that a node in a dyad fulfills a certain purpose in the 
development of software and is hence embedded in a context that supports the 
realization of that purpose. For example, requirements analysis is performed at 
a certain point in time by people possessing the knowledge to select, 
prioritize and validate requirements. Test scenarios may be developed at the 
same time, but require a different set of knowledge in order to realize their 
purpose. When information is aligned between two nodes, the context of the 
nodes differs and hence also the purpose of the information.

In Section~\ref{sec:ProcessMedium} we have motivated how a link between two
nodes can be characterized by a medium. The concept of a medium is however not
able to explain how the information between two nodes is synchronized, i.e.
how the change in purpose is supported by the link. Therefore we use the
concept of mechanism to further characterize the link in information dyads.

To understand the mechanism concept we need to refine our earlier definition of
information as \emph{a coherent collection of related data that is created
during software development, often with a specific purpose in mind}.
Although this definition helps to identify nodes, as discussed in
Section~\ref{sec:ProcessInformation}, it does not provide the granularity to
differentiate between alignment mechanisms. We adopt therefore a definition in
which information has the components of well-formed data \emph{and}
meaning~\cite{floridi_information:_2010}:
\begin{enumerate}
 \item data is well-formed if it has an underlying structure, syntax and
notation
 \item data is meaningful in a certain context, i.e. the meaning of data may
change with its purpose
\end{enumerate}

\begin{example}[t]
 \caption{Step 2.3 - What type of mechanism establishes the
connection?}\label{exa:Example-4}
We start by looking at the dyads that contain both N1, informal requirements,
as an information characteristic in the node. The information in both N2 and N3
is derived, although by different roles, from the informal requirements. The
mechanism for this derivation is in both cases not explicitly specified in
the framework. Hence the connection between the nodes is in both dyads an
\emph{implicit} one. The mechanism in dyad N3-N4 is however a
\emph{transformation} as test sequences are extracted from the testgraph
which are used to animate and test the specification.

Dyads N2-N5 and N3-N6, on the other hand, are explicitly connected by the
tester, creating a mapping between the implementation and the specification
respectively the testgraph. The mere mapping between information in these
dyads does not fulfill the requirements of a transformation mechanism. Consider
for example that the testgraph in N3 is modified due to changes in the informal
requirements. The mapping by itself cannot accommodate such impact but has to
be recreated by the tester.
The mapping identifies corresponding entities in the artifacts, i.e. there is
no change in the notation, excluding therefore also bridge, leading to the
conclusion that we observe a \emph{connection} mechanism.
Dyad N6-N7 is linked again by a \emph{transformation} mechanism since the test
sequences are generated and reflect the information in the testgraph mapping
(N6).
\end{example}

Using these components of information, we can now differentiate between
alignment mechanisms and characterize them according to the means through which
the synchronization and agreement of information, shared between nodes, is
achieved.

\textbf{Transformation:} Information, packaged for one node in the alignment
dyad, is re-packaged in order to satisfy the needs of the other node. A
transformation mechanism that restructures and/or augments the information is
applied, changing the notation and supporting the change in meaning of the data.
Example: A method allows the transformation of a use case into a test model,
changing the notation of the information. The support in adapting the meaning is
given, for example, if relationships to other use cases are pertained in the
transformation and reflected in the model\footnote{Support in adapting the
meaning = preservation of relationships between information across contexts}. We
say that the alignment between nodes is internalized in the
mechanism.

\textbf{Bridge:} Information pertaining to each node is connected and
augmented in order to achieve fitness of purpose in both nodes, changing the
notation. The difference to transformation is that a bridge does not provide
support to adapt the meaning of data within the context change. Example: A
method allows the transformation of use cases into a test model, changing the
notation of information, however without establishing relationships within the
test model that reflect the relationships within the use cases. Adding a new use
case to the test model is supported syntactically, but the positioning in the
test model requires some knowledge which is not provided by the method. We
say that the alignment between nodes is semi-internalized in the mechanism.

\textbf{Connection:} Information pertaining in each node is connected,
establishing a logical link between the two nodes. The mechanism does however
not change the notation, nor does it provide support in adapting the meaning of
the data when changing the context. The difference to the above is that the
connection does not add anything to the information's fitness of purpose, except
establishing a correspondence of the data component of information. Example: A
method allows to link use cases to the corresponding parts of a test model,
without however providing syntactical support. The meaning of the information
within the test model is given only by the connections back to the use cases.
We say that the alignment between nodes is not internalized in the method.

\textbf{Implicit connection:} Information is connected by volatile and
implicit links that are not formalized. Such volatile links can be established
by communication between people or they exist within a shared, commonly agreed
upon, model. As such, it is not evident which of the components of information
are effectively manipulated in a context change.

\vspace{3 mm}
Note that the alignment mechanisms stated above are characterized by their
support in preserving the relationships between information across contexts and
not by their degree of automation. None of the alignment mechanisms implies that
the mechanism is or can be automated.

\subsection{Method context}\label{sec:ProcessContext}
In the previous step we focused on characterizing information dyads in a
rather detailed manner by describing their components. In this step we broaden
our view and study the context in which the described method is embedded.
\citeN{petersen_context_2009} argue that context influences the conclusions
drawn when integrating evidence from industrial studies. In a classification
effort it is hence important to capture the context of the classified objects.
In the following paragraphs we illustrate the context aspects that should be
captured.

\begin{example}[t]
 \caption{Step 3 - In which environment is the method
situated?}\label{exa:Example-5}
The table below summarizes the context of the classified
method~\cite{miller_case_2010}.
\begin{center}
\begin{threeparttable}
\begin{tabular}{lp{11cm}}\toprule
Aspect & Description\\ \midrule
Method setting & implementation of a subset of the GSM specification (\textless 
1 KLOC), focus on functional requirements, model-based testing, bespoke
requirements, natural language requirements and GSM standard specifications \\ 
Focus & 2) Unintentional but noted effect on
alignment\tnote{1}\\
Motivation & None given due to unintentional focus \\
Assumptions & The specification (language) is executable \\
Quality targets & Not stated \\
Validation & Testgraphs are reviewed for correctness and completeness,
testgraph coverage of specification is measured \\
Outcome & Cost-effectiveness comparable to other model-based
techniques, better than manual testing \\
\bottomrule
\end{tabular}
\begin{tablenotes}
 \item [1] Focus is unintentional since their goal was to improve efficiency by
reusing the testgraph in specification \emph{and} implementation testing. The
testgraph concept is interesting from the alignment perspective, since it is
independently derived and hence an alternative representation to formal
specifications of the informal requirements.
\end{tablenotes}
\end{threeparttable}
\end{center}
\end{example}

\textbf{Method setting:} Describe type of development process,
scale\,/\,size (of the project in which the method was applied), focus of
requirements types (functional, quality, both), type of testing (unit,
integration, system, acceptance, formal verification, scenario-based, etc.), and
type of requirements engineering (market-driven or bespoke, use of natural
language primarily or other notation).

\textbf{Focus:} Describe the degree to which alignment of RE and ST is the
primary focus of the method. Is an alignment issue between RE and ST thematized
and addressed (choose 3, 4, or 5)? Are the studied methods/activities embedded
in a software engineering problem that includes, but does not exclusively
discuss RE and ST alignment (choose 1, 2, or 3)?
\begin{enumerate}
 \item Unintentional and undiscussed\,/\,unnoted effect on alignment
 \item Unintentional but noted effect on alignment
 \item Part of purpose was to improve\,/\,affect alignment
 \item Main purpose was to improve\,/\,affect alignment
 \item Intended, main as well as sole purpose
\end{enumerate}

\textbf{Motivating problem:} Describe the driver\,/ intention\,/ motivation to
propose\,/ implement an alignment method.

\textbf{Assumptions:} Describe any constraints or assumptions, e.g. on existing
artifacts or application domains, that the application of the alignment method
makes.

\textbf{Quality targets:} What is aimed to be improved by a better RE and ST
alignment? Examples are reducing time-to-market, test effort, cost,
number of faults, etc. 

\textbf{Validation:} Is there any formal or informal mechanism that supports the
consistency of the shared information? In particular, does the alignment method
provide any support in assessing\,/\,verifying the consistency or correctness of
the shared information?

\textbf{Outcome\,/\,Benefits:} What are the experienced effects of the alignment
method? Note that this should only contain actual (not expected ones) effects
that were established by an evaluation.

\subsection{Dyad structure properties}\label{sec:ProcessStructure}
The central unit of analysis of the REST taxonomy is the information dyad 
(Figure~\ref{fig:InformationDyad}). As we have illustrated in the examples in 
Section~\ref{sec:ProcessInformationDyad}, a REST alignment method may consist 
of several dyads, thus forming a structure that is governed by the components 
of a dyad (Figure~\ref{fig:ExampleDyadStructure}). We have defined a set of six 
properties based upon the characteristics of nodes and links, explained in 
Sections~\ref{sec:PropNumberOfNodes}~-~\ref{sec:PropScope} and illustrated 
in the example for Step 4. The most basic property is the number of nodes 
in a dyad structure. Other properties are derived from the purpose of a node, 
i.e. in which development phase it predominately exists, or the alignment 
mechanism of the link between two nodes. The definition of these properties is 
guided by their usefulness in interpreting and analyzing a dyad structure. In 
Section~\ref{sec:ProcessMethodClassification} we propose a classification of 
alignment methods based upon dyad structure properties.

\subsubsection{Number of nodes (P1)}\label{sec:PropNumberOfNodes}
Links between nodes need to be established and maintained over time. Hence, the
total number of nodes allows one to reason on the (visible, explained)
complexity, and on the effort to establish and maintain REST alignment. A large
number of nodes may indicate a high cost in institutionalizing alignment. 
Furthermore, even tough a larger number of nodes can break down the alignment 
process into manageable sub-tasks, the overall complexity of the method 
increases with the number of nodes, as linking mechanisms between the nodes 
need to be defined and instantiated.

\subsubsection{Branches (P2)}\label{sec:PropBranches}
Looking at an individual dyad, one node acts as a source, the other as a sink 
of information. A branch exists, if the dyad structure is configured such 
that a node acts as a source or sink for more than one node. We provide in 
the example for Step 4 a procedure to identify branches in a dyad structure.

Branches may reduce the complexity of analyzing information (concern 
separation) in sink nodes. However, at the same time branching requires a step 
in which the individually analyzed information is merged, introducing more 
nodes, potentially more effort and an increase of the overall 
methods' complexity.

\begin{example}[t]
 \caption{Step 4 - What is the structure of the 
identified dyads?}\label{exa:Example-6}
The figure below illustrates the dyads that were identified in the method 
presented by \citeN{miller_case_2010}, using the data gathered in the examples 
for Step 2.1, 2.2, and 2.3. 

\begin{center}
  \includegraphics[scale=1.0]{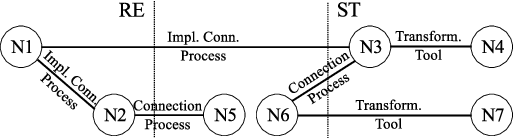}
\end{center}

In this example we show how the dyad structure properties are derived from this 
data.

\emph{P1:} This property is calculated by counting the number of nodes 
identified in the method, which is in this case 7. 

\emph{P2:} The dyads are N1-N2, N1-N3, N3-N4, N2-N5, N3-N6, N6-N7, whereby the 
first node represents the source, and the second node the sink of information. 
From this sequence, we can identify the number of branches by counting the dyad 
instances where a source or sink node occurs more than once. In this example, 
this is the case for 2 source nodes (N1, N3), leading to the conclusion that 
we observe 2 branches in this method.

\emph{P3:} Nodes N5 and N6 are intermediate nodes, hence the value for this 
property is 2.

\emph{P4:} The RE and ST node proportion is 2 (N1,N2) : 3 (N3, N4, N7). 

\emph{P5a/b/c:} This property is extracted by listing the link mechanisms in 
the respective phases. The only Within-RE link is an implicit connection, and 
the Between-Phase links are implicit connection, two connections, and a 
transformation. The only Within-ST link is a transformation.

\emph{P6:} For this property, we look at the nodes that act exclusively 
as source and sink of information in RE and ST respectively. In RE, N1 is the 
only node that acts exclusively as source (N2 is both sink and source). The 
information in N1 is informal requirements, which can be regarded as 
information pertaining to early RE. In ST we have N4 and N7 that act both 
exclusively as sinks. Both contain test sequences (for specification and 
implementation tests respectively), which can be seen as information pertaining 
to late ST. Hence the scope for this method is Early RE - Late ST.
\end{example}

\subsubsection{Intermediate nodes (P3)}\label{sec:PropIntermediateNodes}
Nodes characterized by information that belongs to the design/analysis or
implementation phase of software development are intermediate nodes. Their
existence indicates that the method user is required to have knowledge
outside the RE and ST domain. Intermediate nodes may strengthen overall REST
alignment by integrating analysis/design and implementation, increasing the
traceability. However, intermediate nodes can also imply that the method
may be more invasive to the overall development process.

\subsubsection{RE and ST node proportion (P4)}\label{sec:PropRESTProportion}
Assuming that a node is associated with a certain cost (e.g.
establishing/maintaining the information therein and links in between), it is
of interest to know the node distribution among the RE and ST phases. Such an
evaluation may show which phase is impacted the most by a method. Having
more within-phase nodes (and links) in RE may be beneficial as the level of
abstraction can be adjusted to a level that facilitates the alignment with ST.
On the other hand, nodes (and links) in RE need to be efficient as requirements
may change and may be refined continuously, promoting less nodes in the RE
phase.

\subsubsection{Within-RE (P5a) / Between-Phase (P5b) / Within-ST links
(P5c)}\label{sec:PropPhaseLinks}
Based upon the information characterizing a node, we can approximate roughly its
primary development phase and whether it is located early or late in that phase.
This allows us to reason upon the linking mechanisms within the RE and ST
phases, and between those phases. The reason for such a distinction emerges from
the assumption that each phase has different properties that need to be taken
into account by the applied linking mechanism(s). For example, Within-RE links
may need to accommodate frequent requirement changes, informally specified
requirements and different requirement abstraction 
levels~\cite{gorschek_requirements_2006}. Within-ST links typically link
test cases on different abstraction levels, whereas Between-Phase links require 
a more complex mapping since the context in the phases differs.

\subsubsection{Scope (P6)}\label{sec:PropScope}
By approximating the location of nodes in development phases, we can
distinguish between early and late nodes. The distinction between early and 
late requirements is often made to differentiate between an ``understanding'' 
and ``describing'' phase in RE (e.g. in the Tropos development methodology 
~\cite{mylopoulos_tropos:_2000}). Similarly, one can also distinguish between 
early and late phases in ST.  For example in RE, early artifacts can be natural 
language requirements and use case descriptions, whereas requirements models 
can be put closer to the Analysis/Design phase.
Similarly, test scenarios, abstract test cases and test plans can put on the
left, executable test cases on the right spectrum in ST.
This allows us to reason upon the scope of an alignment method with respect
to the RE and ST phases, and its implications. For example, a method may not
provide a link between natural language requirements and more formalized 
models. In a scenario where such a link would be beneficial, the method may 
need to be extended or combined with other approaches.

\subsection{Method classification}\label{sec:ProcessMethodClassification}
Up until now we have illustrated the REST taxonomy from the viewpoint of a 
single alignment method, that is, describing the process of identifying 
information dyads, extracting the context in which the method is applied/used, 
and characterizing the method through dyad structure properties. In this section 
we expand this view by proposing a classification schema for alignment methods, 
based upon dyad structure properties.

\subsubsection{Overview of the classified methods}
We have applied the taxonomy, in total, on 13 alignment methods. In the 
remainder of this paper they are referenced as cases A-M:
A~\cite{guldali_torc:_2011},
B~\cite{flammini_automatic_2009}, C~\cite{de_santiago_junior_generating_2012},
D~\cite{el-attar_developing_2010}, E~\cite{miller_case_2010},
F~\cite{nebut_automatic_2006}, G~\cite{conrad_systematic_2005},
H~\cite{abbors_tracing_2009}, I~\cite{damian_requirements_2005},
J~\cite{arnold_scenario-based_2010}, K~\cite{zou_control_2008},
L~\cite{siegl_model_2010}, and M~\cite{metsa_testing_2007}.

Cases F-M stem from the set of papers that were used for taxonomy construction, 
whereas cases A-E stem from a search in literature, as explained in
Section~\ref{sec:TaxonomyConstruction}. Since the identification and 
characterization of information dyads is a crucial step in the application of 
the taxonomy, we provide additional examples of this process on cases A-D in 
Appendix~\ref{sec:AppendixTaxAppMethodResults} (case E has served as a running 
example throughout this section).

\subsubsection{Classification schema}\label{sec:ClassificationSchema}
The schema we adopt aims at providing a meaningful and useful classification of 
alignment methods. Looking at the definitions of the dyad structure properties 
in Section~\ref{sec:ProcessStructure}, we can observe that properties P1, P2, 
P3 and P5 characterize the complexity, and P4 and P6 describe the focus and 
scope of the method. We chose therefore a simple two-dimensional schema that 
encodes the overall complexity of the classified method on the vertical and the 
focus/scope on the horizontal axis.
Since we use multiple properties to represent complexity, we define the 
following order for sorting a method on the complexity dimension: 
\begin{enumerate}
 \item P1 (Number of nodes): this is the main sorting criterion as each node, 
through its associated information, contributes to the need of maintaining 
consistency (otherwise, the very purpose of the method would be violated).
 \item P5b (Between-Phase links): each link that crosses a phase boundary 
(e.g. from RE to design, RE to ST), contributes to the overall complexity as 
information from different contexts is linked. We use the number of 
Between-Phase links as the second sorting criterion.
 \item P5ac (Within-RE and Within-ST links): linking information on different 
abstraction levels is less involved than linking information from different 
contexts; as these links lie within one phase, we use the number of 
Between-Phase links as third sorting criterion.
 \item P2 (Branches): even though related to the number of links, branches are 
less indicative for the overall method complexity as they act locally (i.e. 
within a dyad) as an agent to reduce complexity. They are therefore the fourth 
sorting criterion.
 \item P3 (Intermediate nodes): everything else being equal, intermediate nodes 
are the fifth sorting criterion.
\end{enumerate}

\begin{figure}
 \begin{center}
  \includegraphics[scale=1.0]{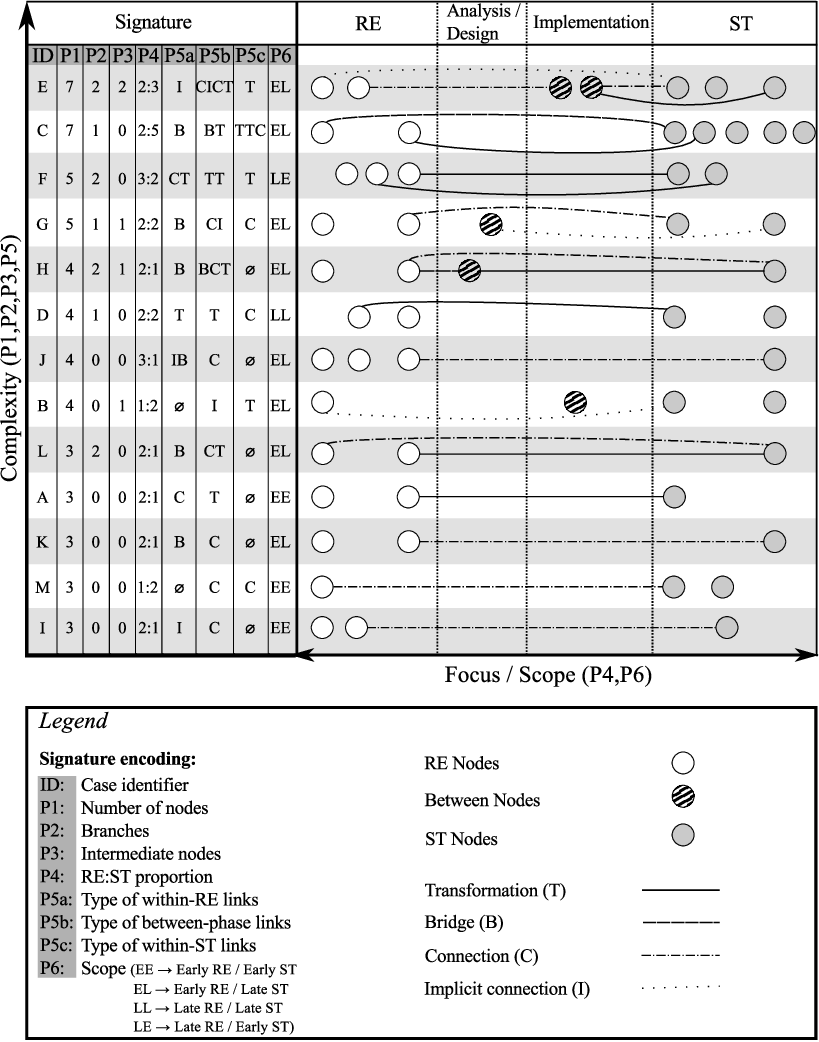}
\end{center}
\caption{Cases classified in the REST taxonomy}
\label{fig:NodeDistribution} 
\end{figure}

Note that the ordering above considers only complexity defined by the 
properties we have identified. For example, we do not classify the information 
in a node itself. Hence, the complexity of an alignment method, as defined by 
the classification schema, is an approximation that can be improved by a 
more fine-grained characterization of a node's information component. As a 
consequence, the presented classification does \emph{not} provide any statement 
on the performance of the classified alignment methods. Nevertheless, the 
qualitative classification of the method context (see 
Section~\ref{sec:ProcessContext} and Table~\ref{tab:CaseContext}), in particular 
the method setting, assumptions and quality target aspects, provide means to 
interpret and judge the quantitative classification.

The second dimension of the schema (horizontal axis), characterizes the methods 
according to their focus (P4) and scope (P6).

\subsubsection{Classification results}
Figure~\ref{fig:NodeDistribution} shows the 13 classified cases. In the 
left-most column, we encode the dyad structure details in a signature, whereas 
in the right part of the figure, the structure is represented graphically 
(only Between-Phase links are drawn). Note that cases A, K, M and I 
have the same complexity according to the sorting criteria defined in 
Section~\ref{sec:ClassificationSchema}. We analyze the results of the 
classification in Section~\ref{sec:AlignmentAsActivity}. 

\section{Taxonomy construction and validation method}\label{sec:Validation}
In this section we describe how we constructed and validated the taxonomy, and 
discuss threats to validity of this approach.

\subsection{Iterative construction and
validation}\label{sec:TaxonomyConstruction}
In Section~\ref{sec:DevelopingTaxonomies} we motivated why the taxonomy was
constructed in a bottom-up fashion. We started by sampling alignment
methods published in literature. The initial sample consisted of 16
publications that were analyzed in a systematic mapping study on aligning
requirements specification and testing~\cite{barmi_alignment_2011}. Since the
mapping study had limitations, as further discussed in
Section~\ref{sec:InternalValidity}, we added 10 more publications that we
regarded relevant by reading title and abstract. Hence, the taxonomy 
construction pool consisted of 26 publications, from which 15 were used in the 
taxonomy construction process (iteration 1-4). Although we used all 15 
publications in the construction process, we classified 8 of them (cases 
F-M) and excluded 7 for the following reasons:
\begin{itemize}
 \item the publication covered only the RE aspect, leading to the decision 
that the method, described in this particular study, is out of scope: 4 
publications 
(\cite{hayes_advancing_2006,grunske_specification_2008,%
mugridge_managing_2008,niu_aspects_2009})
 \item the publication only sketched a solution proposal or reported lessons 
learned, and was therefore not descriptive enough to warrant a classification: 
2 publications (\cite{winbladh_automated_2006,kukkanen_applying_2009})
 \item the publication is a predecessor to a publication that has been 
classified in this paper: 1 publication (\cite{nebut_requirement-based_2004})
\end{itemize}

In iteration 5 we classified 5 more publications (cases A-E) that were not 
included in our initial pool, resulting in a total of 13 classified methods that 
are presented in this paper. 

Three researchers were involved at different stages in the construction and 
validation of the taxonomy. The milestones of this iterative process are 
illustrated in Figure~\ref{fig:Timeline}. We discuss these iterations in the 
following subsections.

\subsubsection{Iteration 1}\label{sec:Iteration1}
\begin{figure}
 \begin{center}
 \includegraphics[scale=0.66]{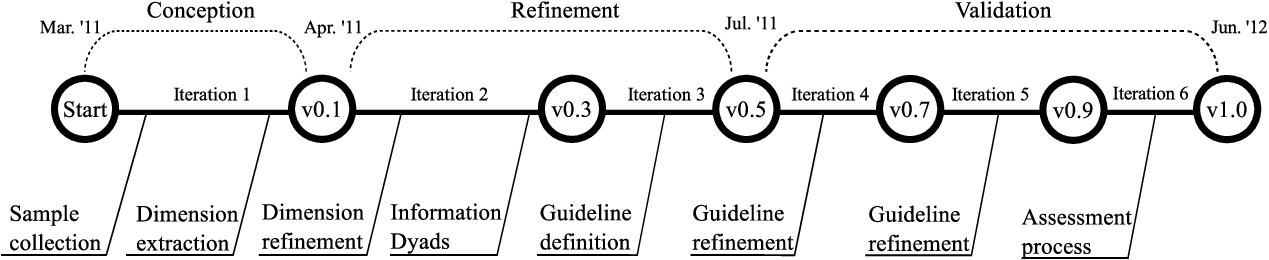}
\end{center}
\caption{REST Taxonomy construction and validation process}
\label{fig:Timeline} 
\end{figure}

In the first iteration we chose five publications to bootstrap a set of
dimensions for the classification of alignment methods. The first author applied
open coding~\cite{robson_real_2002} on each method description and consolidated
the emerged codes into dimensions characterizing the alignment approaches (v0.1
of the taxonomy, see Figure~\ref{fig:Timeline}).

The strategy was to identify commonalities or distinguishing aspects in the
described methods. For example, one common aspect was that information from the
requirements engineering phase is reused in downstream development and
eventually in system or acceptance testing, leading to a dimension describing
information source and sink. Another early dimension, describing the packaging
of information (e.g. in natural language, diagrams, etc.), characterized whether
information is used ``as is'' or if it is adapted or enriched for downstream
use.

\subsubsection{Iteration 2}\label{sec:Iteration2}
In the second iteration, the second author was invited to verify whether the
identified dimensions characterize the methods in an useful manner. We re-used
the publications from iteration 1. Although the definitions of the dimensions
were refined in this iteration, we realized that a characterization of the
heterogeneous set of methods would not be possible with static dimensions
describing the method as a whole. For example, methods could be characterized by
several information sources and sinks. Hence we introduced the concept of
information dyads which allowed us a more fine-grained and flexible
characterization, leading to v0.3 of the taxonomy.

\subsubsection{Iteration 3}\label{sec:Iteration3}
The first and second author chose five new publications for the third iteration.
The dimensions, now consolidated in the information dyad construct
and the context aspects, were further refined and a guideline was developed 
(v0.5 in Figure~\ref{fig:Timeline}). We chose two additional publications and 
exemplified the application of the taxonomy in the guidelines.

\subsubsection{Iteration 4}\label{sec:Iteration4}
We invited the third author to validate the updated taxonomy and the operational
guidelines developed in the previous iteration. We chose three new methods
from the sample set and all three authors independently applied the taxonomy.
We analyzed the results in a post-mortem.

On method 1, we achieved in general a good agreement, having however some
variance on the identified medium (link characteristic) and the level of focus
on alignment (context aspect). Looking at the guidelines, we identified
the definitions of the different media and alignment focus levels as a cause
for the disagreement and clarified them. On method 2 and 3 we observed a larger
variance among the three analysts. The major reason was a disagreement on
whether the method is in the scope of the taxonomy, i.e. if it can be
classified as an alignment method. Hence we added Step 1, identifying the
relevance of the studied method, in the taxonomy application process (see
Table~\ref{tab:RESTProcess}).
The intention of the scope criterion is to clarify that we are interested in
classifying methods that consider both requirements engineering and testing
aspects. Methods that bridge other gaps, e.g. between design and test, are by
this definition excluded.

\subsubsection{Iteration 5}\label{sec:Iteration5}
The aim of this iteration was to apply the taxonomy on a set of methods that
were not included in the initial set of publications. To this end, we chose four
premium venues for publications in Requirements Engineering (Requirements
Engineering Journal), Verification\,\&\,Validation (Software Testing,
Verification and Reliability) and software engineering in general (IEEE
Transactions on Software Engineering, Software Quality Journal)
as the population for drawing our sample. We chose 2007 as the starting point
for our search since we did not aim to perform a systematic literature
review~\cite{kitchenham_guidelines_2007} and hence do not claim complete
time coverage. Furthermore, 2007 seemed to be a good starting point since
\citeN{cheng_research_2007} and \citeN{bertolino_software_2007} called
for a closer collaboration between requirements engineering and software testing
research at FoSE that year. 

The first author manually searched, by reading title and abstract, 635
publications from the period 2007\,-\,2011, applying the criteria defined in
Section~\ref{sec:ProcessRelevance}. After applying the scope criterion on the
abstracts, 148 publications remained for full text screening. In this step, the
scope criterion was applied a second time, excluding methods that were only
partially bridging the gap between RE and ST, e.g. verification of UML
models~\cite{siveroni_uml-based_2010}, derivation of specifications from
requirements~\cite{seater_requirement_2007}, or derivation of test cases from
design artifacts~\cite{pickin_test_2007}, leading to 24
publications\footnote{In the title and abstract screening we were rather
inclusive, resulting in many irrelevant studies in the set for full text
reading.}. On these, we applied the comprehensiveness criterion, including only
those methods for which we could answer the questions posed in steps 2 and 3 of 
the taxonomy application process (see Table~\ref{tab:RESTProcess}). We
concluded the search with 5 publications describing alignment methods and 
applied the taxonomy, leading to two further refinements to the guidelines:
\begin{longitem}
 \item Introduction of use-relationships between nodes. For example in Case B
(see Table~\ref{subtab:case_B}), node N3 contains information that is necessary
for the method, is however not related with any other node through a link
mechanism. The use-relationship legitimates N3 in the dyad structure, increasing
the richness of the method characterization.
 \item Introduction of a further aspect in context identification (Step 3) of
the taxonomy process (Section~\ref{sec:ProcessContext}). Recording assumptions
or constrains helps to understand under which circumstances a method may be
applicable.
\end{longitem}
The results of the taxonomy application on four cases (one method has served as
the running example in Section~\ref{sec:TheTaxonomy}) are illustrated in
Appendix~\ref{sec:AppendixTaxAppMethodResults}.

\subsubsection{Iteration 6}\label{sec:Iteration6}
The aim of this iteration was to evaluate whether the REST taxonomy provides
support in identifying misalignment in a development organization. The first
author developed an assessment guideline and procedure, REST-bench, that
is powered by the concepts underlying the REST taxonomy. The approach and
assessment results are described in Section~\ref{sec:AlignmentAsState}.

\subsection{Validity threats}\label{sec:ValidityThreats}
The bottom-up construction of the taxonomy is subject to several validity
threats~\cite{wohlin_experimentation_2000}. 

\subsubsection{Internal validity}\label{sec:InternalValidity}
The systematic mapping study by \citeN{barmi_alignment_2011},
from which we sampled publications and bootstrapped the dimensions of the
taxonomy, was initially designed to identify alignment methods focusing on
\emph{non}-functional requirements and software test. Although the scope has
been extended to include also functional requirements, the mapping study may
have missed relevant studies\footnote{Two studies
(\cite{flammini_automatic_2009,el-attar_developing_2010}) that were identified
in the manual search during the validation were not identified by the
search (November 2010) in the mapping study.}. We added therefore 10 studies
that we considered relevant. Still there is a moderate threat that our sample of
methods was biased. 

\subsubsection{Construct validity}\label{sec:ConstructValidity}
The identification of characteristics defining an alignment method, as
described in Section~\ref{sec:TaxonomyConstruction}, is subject to mono-method
bias~\cite{wohlin_experimentation_2000}. The first author performed the initial
analysis and may have subjectively biased the taxonomy construction. To
counteract this threat, we designed the taxonomy construction as an iterative
process, involving multiple researchers with expertise in both requirements
engineering and software verification\,\&\,validation.

\subsubsection{External validity}\label{sec:ExternalValidity}
During the validation we performed a manual search on four
premium journals, identifying further methods and applying the taxonomy. The
selection was based on reading the title and abstract of the study, searching
for indications that both requirements engineering and software testing aspects
were discussed. This means that ``partial'' solutions that bridge for example
the gap between user requirements and requirements specifications
(e.g.~\cite{liu_integrating_2009}), requirements to design
(e.g.~\cite{valderas_introducing_2009}), or design to test
(e.g.~\cite{samuel_automatic_2007}) were not considered to validate the
taxonomy.

The goal was to validate whether the taxonomy can be applied on
alignment methods that were not part of the construction sample and not to
identify and classify all existing methods. A thorough overview of alignment
methods could be performed by conducting a systematic literature
review~\cite{kitchenham_guidelines_2007}. The review could be designed to
include the type of the above mentioned solutions, and, by using the taxonomy
presented in this paper as an analysis aid, provide practitioners support in
selecting and combining methods, as well as provide researchers an overview for
further empirical or conceptual research.

\section{Method evaluation using the REST
taxonomy}\label{sec:AlignmentAsActivity}
In this section we elaborate on the application of the taxonomy, exemplifying
analysis on two levels. First, we show the potential of the taxonomy as a mean
to describe the state-of-the-art of REST alignment methods in
Section~\ref{sec:AASummaryAnalysis}. Then, in
Section~\ref{sec:AADyadStructureAnalysis} we illustrate the application of the
dyad structure property analysis introduced in
Section~\ref{sec:ProcessStructure}.

\subsection{Summary analysis}\label{sec:AASummaryAnalysis}
In Figure~\ref{fig:NodeDistribution} we have classified the alignment methods 
presented in the 13 studied cases which allows us to perform basic quantitative 
analysis. We observe that the mode for number of dyads is 2, the median is 3. 
This indicates that methods with more than 4 dyads are uncommon. A similar
observation can be made on the number of nodes, with a mode of 3 and a
median of 4. Methods with more than 4 nodes are not common.

The right part of Figure~\ref{fig:NodeDistribution} shows the distribution of
nodes in the respective software development phases. The links between nodes
highlight dyads which span over distinct development phases. Overall, we can
observe a slight majority of nodes in the earlier phases (RE:26, ST:24). This
tendency is more pronounced (RE:17, ST:12) if we exclude the cases C, E, F and 
G, which have an untypical (w.r.t. the mode) high number of nodes.

Looking at the alignment mechanisms, \emph{connection} and 
\emph{transformation} are the most common alignment mechanisms with a frequency 
of 15, followed by \emph{bridge} (9) and \emph{implicit connection} (6). The 
proportion of within and between phase links is 1:1, i.e. there are 22
links between and equal as many links within development phases.
Figure~\ref{fig:NodeDistribution} illustrates also the types of mechanisms
linking nodes in distinct development phases (within-phase links are not shown).
Overall, we observe that for the \emph{connection} mechanism the between-phase
links dominate (9 out of 15), whereas for the \emph{bridge} mechanism
within-phase links dominate (7 out of 9). For the \emph{transformation} and
\emph{implicit connection} mechanism, within- and between-phase links are
equally distributed (7 within, 8 between and 3 within, 3 between). The
occurrences of alignment medium are as follows: \emph{Process} (22),
\emph{Tool} (17), \emph{Structured artifact} (4) and \emph{Organization of work
environment} (1).

The \emph{connection} mechanism, which we defined as establishing a logical
link between information in two nodes (see Section~\ref{sec:ProcessMechanism}),
can be viewed as a mean to establish traceability. Given that this alignment
mechanism, together with \emph{transformation}, was observed most frequently, 
we can assert that establishing traceability is, in general, a main concern of 
the studied alignment methods. 
As shown in the analysis, the between-phase links with a \emph{connection} type
mechanism dominate, mapping for example technical requirements to test scenarios
(Case I~\cite{damian_requirements_2005}), requirements classification trees to
logical test scenarios (Case G~\cite{conrad_systematic_2005}), or test reports
to requirements models (Case H~\cite{abbors_tracing_2009}). This observation
concurs with Gotel and Finkelsteins'~\citeyear{gotel_analysis_1994} definition
of requirements traceability referring to ``\emph{the ability to describe and
follow the life of a requirement, in both forwards and backwards direction}''.
Note however that traceability (respectively nodes) to the analysis/design and
implementation phase is sparse due to our selection criteria for RE and ST
alignment methods (we excluded methods which addressed only a subset of the
development phases). One exception is Case E~\cite{miller_case_2010} in which
formal specifications and testgraphs are mapped to the implementation.

In the analysis we have identified eight between-phase links featuring a
\emph{transformation} mechanism. Looking at Figure~\ref{fig:NodeDistribution},
the corresponding nodes are almost exclusively (except Case
D~\cite{el-attar_developing_2010}) located in the late RE phase, preceded by one
or more nodes. This pattern is expected for model transformations, e.g. as in
Case F~\cite{nebut_automatic_2006} (use case transitions system
$\rightarrow$ test objectives) or Case L~\cite{siegl_model_2010} (time usage
model $\rightarrow$ test cases). It also shows that transformation links from
early RE phases to ST are not common.

One aspect that is currently not considered in the taxonomy is the cost of
creating and maintaining the links between nodes and hence maintaining the
alignment. Would the taxonomy have been available to the originators of the
discussed alignment methods, they could have assigned a relative cost to each
link. That would allow us to compare the cost of the methods in the distinct
software development phases. Furthermore, an absolute cost measure would allow
one to reason on return on investment~\cite{unterkalmsteiner_evaluation_2012},
provided that the benefits can be estimated too.

\subsection{Dyad structure analysis}\label{sec:AADyadStructureAnalysis}
\begin{table}[t]
\tbl{Example of trade-off analysis using dyad structure 
properties\label{tab:TradeoffAnalysis}}{%
 \begin{tabular}{lp{2.6cm}p{6cm}p{6cm}}
  \toprule
  Prop.{$^a$} & Value{$^b$} & Benefit & Liability\\ 
  \toprule
  \multicolumn{4}{c}{Case A}\\
   \midrule
   P1 & 3 & Few artifact types involved & Transf. in dyad N2-N3 complex
and iterative \\
   P4 & 2:1 & Reduces ST effort & Limited to abstract test cases\\
   P5a & Connection (N1-N2) & Efficient for new/changed requirements
& None\\
   P5b & Transformation (N2-N3) & Defined and repeatable process & Relies on
specific notation for requirements \\
   P6 & Early RE - Early ST & Supports ST in defining test scope & Concrete test
cases are not created\\
   \bottomrule
\\
   \multicolumn{4}{c}{Case B}\\
   \midrule
   P1 & 4 & No new artifact types are introduced & Tailored for a specific
reference architecture \\
   P3 & 1 & Supports the semi-automated generation of test cases & Incorrect
system configuration may cause faulty executable tests\\
   P4 & 1:2 & None & Abstraction level not easily matched\\
   P5b & Implicit connection (N1-N2) & Given natural language requirements
(NLR's) are appropriately formulated, mapping to abstract test cases is
straightforward & Mapping is not explicit; domain knowledge required to create
mapping\\
   P5c & Transformation (N2-N4) & Instantiation of abstract test cases for a
specific configuration & Correctness of configuration itself is not verified\\
   P6 & Early RE - Late ST & Tests cover requirements considering specific
configurations & Early link (N1-N2) does not address different abstraction
levels of NLR's and test cases \\
   \bottomrule
\\
   \multicolumn{4}{c}{Case C}\\
   \midrule
   P1 & 7 & Broken down complexity into simple steps & Artifacts needed solely
in testing \\
   P2 & 1 & Separation of concerns (Scenario development / Statechart model) &
Information needs to be merged again for testing purpose\\
   P4 & 2:5 & Analysis of reqs. tailored to support testing & Limits reuse in
other development phases\\
   P5a & Bridge (N1-N3) & Enables transformation for between-phase
link (N3-N5) & Domain knowledge required to establish and maintain\\
   P5b & Bridge (N1-N2) / Transformation (N3-N5) & Formalized
and automated transformation & Transformation depends on three previous links \\
   P5c & Connection (N2-N4) / Transformation (N5-N6, N6-N7) &
Step-wise refinement and adaption of abstraction level... & ...except for N2-N4,
which may introduce a bottleneck when scenarios or SRSs change\\
   P6 & Early RE - Late ST & Enables traceability, allowing to verify
requirements coverage & Although partly automated overall, nodes in early RE are
linked manually\\
   \bottomrule
\\
   \multicolumn{4}{c}{Case D}\\
   \midrule
   P1 & 4 & Few newly introduced artifact types & None \\
   P2 & 1 & Enables link between problem and solution domain & Needs to be
maintained in parallel as requirements change to avoid inconsistencies\\
   P4 & 2:2 & Similar abstraction level in both RE and ST & None\\
   P5a & Transformation (N1-N3) & Defined and structured process & Requires
training to apply correctly\\
   P5b & Transformation (N1-N2) & Usable even without executable test cases &
Uses information from different models, potentially causing inconsistencies \\
   P5c & Connection (N2-N4) & Enables traceability & None\\
   P6 & Late RE - Late ST & Focus on artifacts that have similar abstraction
level & Does not cover early RE, e.g. natural language requirements
specifications \\
   \bottomrule
  \end{tabular}}
  \begin{tabnote}
   \tabnoteentry{$^a$}{The abbreviations in this column refer to
the dyad structure properties defined in
Section~\ref{sec:ProcessStructure}: P1 (Number of nodes), P2 (Branches),
P3 (Intermediate nodes), P4 (RE and ST nodes proportion), P5a (Within-RE links),
P5b (Between-Phase links), P5c (Within-ST links), P6 (Scope).}
   \tabnoteentry{$^b$}{The values in this column are based on the results of
the taxonomy application illustrated in
Appendix~\ref{sec:AppendixTaxAppMethodResults}.}
  \end{tabnote}
\end{table}

The goal of this analysis is to provide means to reason on the benefits and
liabilities of REST alignment methods. In particular, the analysis allows to
discuss the trade-offs of methods on a level that is relevant for practitioners
that seek to adopt a method and to improve REST alignment in their context.
The trade-off analysis is based upon the dyad structure properties defined in
Section~\ref{sec:ProcessStructure}.

For each of the properties, a value can be extracted from the the dyad
structure that has been established when applying the taxonomy on the REST
alignment method. Then, benefits and liabilities can be elaborated for each dyad
structure property. Table~\ref{tab:TradeoffAnalysis} illustrates this analysis
on four methods, using the results from the taxonomy application shown in
Section~\ref{sec:ProcessMethodClassification}.

The current set of dyad structure properties defines four properties that can,
by their nature, be found in every REST alignment method: each method consists
of two or more nodes (number of nodes (P1)), of which one or more nodes belong
either to the RE or ST development phases (RE and ST nodes proportion (P4),
between-phase links (P5b) and scope (P6)). As such, these four properties
underline the scope criterion of alignment methods described in
Section~\ref{sec:ProcessRelevance} and hence define a minimum set of properties
for a REST alignment method. 

The remaining properties (branches (P2), intermediate nodes (P3), within-RE and
within-ST links (P5a, P5c)) are not featured by every alignment method, as seen
for example in Case A in Table~\ref{tab:TradeoffAnalysis}. They do however
provide relevant information on the alignment methods as the benefits
and liabilities show in Cases B, C and D. Concluding on the dyad structure
analysis, the six properties provide means to characterize and analyze
individual REST alignment methods, are however not adequate to enable a
comparison between methods as not all properties can be observed in every
method. The assessment of benefits and liabilities in
Table~\ref{tab:TradeoffAnalysis} should therefore be interpreted in the context
of the respective methods. For example, the methods presented in Cases A and B, 
with a relatively low complexity according to our classification, rely on a 
certain requirements specification form and reference architecture (see 
assumptions in Table~\ref{tab:CaseContext}). Furthermore, the motivations and 
targeted goals of these methods differ (test process efficiency vs. test 
coverage), such that general conclusions on the adequacy of a method, based 
alone on the quantitative classification of dyad structure properties, are 
likely not to be accurate. In order to reduce the risk of a misleading 
taxonomy application, we recommend therefore to interpret the quantitative 
classification in conjunction with the qualitative classification (method 
context), which provides information that indeed allows adequacy judgments on a 
method with respect to particular company settings and goals.

\subsection{Lessons learned and limitations}
In Iteration 5 of the taxonomy construction (see
Section~\ref{sec:TaxonomyConstruction}) we searched in 635 publications for REST
alignment methods. We expected to identify a number of publications
that would allow us to illustrate the characteristics of the
state-of-the-art REST alignment methods. We excluded however, applying the scope
criterion (see Section~\ref{sec:ProcessRelevance}) 630 publications, indicating
that there is a lack of research and solution proposals on supporting the
alignment between RE and ST. On the other hand, we identified several
``partial'' solutions (from the RE and ST alignment perspective), that address
specific gaps. From the RE perspective we observed efforts to improve
the traceability from requirements engineering activities and artifacts to
design
(e.g.~\cite{houmb_eliciting_2010,navarro_semantic_2010,pires_integrating_2011}),
 and similarly, from the ST perspective, test generation from design artifacts
(e.g.~\cite{xu_testing_2010,kundu_system_2009,pickin_test_2007}). Together
with the low number of identified RE and ST alignment methods, this indicates
that the envisioned closer collaboration between RE and ST researchers
~\cite{cheng_research_2007,bertolino_software_2007} is still in its early
development, that there is potential in streamlining the efforts in the
respective areas, and that the proposed taxonomy can indicate gaps in research.
For example, it could be investigated whether the partial solutions can be
combined and which adaptions need to be made to construct new REST alignment
methods.

Regarding the components of the taxonomy, we experienced that the classification
of medium, characterizing the link in an information dyad, can be confounded.
The medium characterizes a link between nodes, not the information in the node.
This makes the analysis conceptually more difficult and may lead the analyst to
(wrongly) classify the medium of information in the node instead of the link. On
the other hand, applying the taxonomy according to the guidelines
(Section~\ref{sec:TheTaxonomy}) and limiting the characterization of the medium
on the link, leads to classifications were the medium is often a process (i.e.
the process/activity transferring the information from node A to node B). 
A factor that contributes to the difficulty in classifying the link medium is
that the taxonomy defines a medium both as a carrier of information and also as
a facilitator that enables information transfer (see
Section~\ref{sec:ProcessMedium}). Further use or application of the taxonomy
might show whether medium as a characteristic of an information link needs to
be refined, either by a more precise definition or by modeling it in a
different manner. In this study we have tried to strike a balance between
analytical depth and taxonomy usability and thus opted for not refining the
concept of a medium.

The construction and application of this particular taxonomy was subject to a
circular problem. The publications and RE and ST alignment methods we studied
were likely not written with the concept of an information dyad in mind. Still
the concept can be used to characterize a wide variety of alignment methods.
Extracting the characteristics is for the same reason challenging and for some
cases not objectively possible e.g.~\citeN{de_caso_automated_2010},
\citeN{uzuncaova_incremental_2010}, and \citeN{grieskamp_model-based_2011},
which we excluded from further analysis in Iteration 5 although seemingly
relevant. Would there have been a taxonomy on RE and ST alignment methods
available when these methods were conceived, they may have been reported
differently. We propose that our taxonomy can be used to structure and give
detail in future papers that report on alignment methods. Such an effect has
been observed in the Global Software Engineering community after the publication
of a classification scheme for empirical research in the
area~\cite{smite_reporting_2008}.

\section{Industrial case study using REST-bench}\label{sec:AlignmentAsState}
To make the REST taxonomy relevant for industrial assessment of alignment we 
created a lightweight framework. REST-bench is powered by the REST taxonomy, 
reusing the information dyad and dyad structure concepts presented in 
Section~\ref{sec:TheTaxonomy}, but also includes process elements (how to use 
REST-bench) and analysis and visualization elements. This section shows the 
industrial application and test of the taxonomy through the use of REST-bench in 
an case study. We describe REST-bench in Section~\ref{sec:REST-benchMethod}, 
present the results in
Section~\ref{sec:ASResults} and illustrate the dyad structure analysis in
Section~\ref{sec:ASDyadStructureAnalysis}.

\subsection{REST-bench process overview}\label{sec:REST-benchMethod}
The goal of the assessment is to identify improvement opportunities in the
coordination between the requirements engineering and the system testing
organization. In order to elicit information on the current state of affairs,
REST-bench focuses on the relationships between artifacts created
and used by the different roles in the software organization, particularly by RE
and ST roles. The choice of centering the assessment around artifacts is
motivated by their importance in carrying information, which is the basis for
characterizing alignment (as we have illustrated with the information dyad in 
the REST taxonomy). The objectives of the assessment are to:
\begin{itemize}
 \item elicit, from the RE and the ST perspective, the artifacts that they
create and for which purpose these artifacts are created
 \item contra-pose those two perspectives to identify disagreement 
 \item identify deficiencies in the creation/use of artifacts that impede
alignment
\end{itemize}

The procedure to achieve these objectives is summarized in the following steps.
\begin{longitem}
 \item [STEP 1 (SELECTION):] Representatives from the RE and the ST role are
interviewed. One important constraint for the selection of interviewees is that
they have or are currently collaborating in the same project. This allows
elicitation of information on the actually created and used artifacts
instead of referring to what is prescribed or recommended by the official
process in an organization. The interviews with the RE and ST representative are
conducted separately, following a guideline that supports the analyst to collect
information regarding the context of the agreed upon project and the artifacts
created and used in the project. 
 \item [STEP 2 (MAP CREATION):] The analyst creates an artifact map which shows
use-relationships between the artifacts in the studied project. Furthermore,
each artifact is annotated with the role(s) that created and used it during the
project. This artifact map merges the perspectives of the RE and ST
representatives, providing a basis for discussion and analysis during the
following step.
 \item [STEP 3 (ANALYSIS WORKSHOP):] The analyst, RE and ST representatives
conduct a workshop in which the artifact map is reviewed. Artifacts,
relationships, users and creators are confirmed, modified or extended. The RE
and ST views are merged and the analyst uses the dyad structure properties to
elaborate together with the workshop participants potential improvements.
\end{longitem}

We reuse the concept of dyad structure properties, introduced in 
Section~\ref{sec:ProcessStructure}, in the analysis of the nodes 
represented in the artifact map, refining however the definition of the 
properties to the particular context of assessing alignment 
(alignment-as-state). For each property, we propose a set of questions the 
analyst may ask during the workshop to initiate the discussion and analysis.  

\subsubsection{Number of nodes (P1)}
In the assessment of an organization, the number of nodes relevant for REST
alignment is a first indicator for identifying bottlenecks or overhead.
Too few nodes can indicate challenges in coordinating RE and ST activities 
since the necessary information is not shared effectively. On the other hand, 
too many nodes can indicate that much effort is spent on keeping these nodes 
up-to-date and synchronized, not to mention roles and responsibilities. Each 
node can represent an individual, department or role in an organization. If 
this is the case, each link between nodes can imply the creation of overhead 
and possibilities for everything from 
misunderstandings~\cite{gorschek_requirements_2008} to miscommunications due to
sub-optimization~\cite{fricker_handshaking_2010}.
\begin{itemize}
 \item Is there an information need that was not fulfilled by the used 
artifacts?
 \item (If applicable) Given that artifact X doesn't have any user / is only 
used by A, could the information in artifact X be merged into artifact Y?
\end{itemize}

\subsubsection{Branches (P2)}
The contribution of branches to REST alignment needs to be evaluated. In 
particular, whether the information in a branch node is actually used in either 
a RE or ST activity, or whether the branch has a different purpose. Such a 
distinction is needed for a more accurate estimation of the spent effort for 
REST alignment. In extreme cases a branch can be seen as ``dead'', adding 
nothing to alignment, which becomes visible when applying the taxonomy 
evaluation, but is not obvious in every-day activities.
For example, test plans that are derived from initial requirements
specifications have little value if not updated and maintained as the
development proceeds and requirements change. In such a scenario a
detailed test plan, e.g. specifying which requirements are covered by which test
cases, is waste as it won't be used, due to its inaccuracy, during testing.
\begin{itemize}
 \item How is the information in artifact X kept consistent with the 
information in artifact Y, in the case Z changes (Z has two links, a branch, to 
X and Y)?
 \item If inconsistencies between artifacts X and Y arise, how does that impact 
the users of those artifacts and their work?
\end{itemize}

\subsubsection{Intermediate nodes (P3)}
It is important to identify and to understand the purpose of intermediate nodes 
for two reasons. First, changes to nodes in the design or implementation phase 
may inadvertently affect REST alignment. For example, the replacement of a 
design artifact with more frequent meetings between analysts and programmers 
may reduce the documentation effort, however breaking at the same time an 
important link that establishes a connection between high-level requirements 
and system tests. This is especially relevant for organizations moving from a 
plan-driven to a lean/agile development process where an improvement in saving 
of perceived overhead in terms of documentation can be a sub-optimization from a
product perspective~\cite{gorschek_requirements_2008}. Second, intermediate
nodes may also represent an overhead that may be eliminated without affecting
REST alignment. Both scenarios become visible through the application of a REST 
alignment analysis.
\begin{itemize}
 \item Do the creators of artifact X (designers), deliver timely, i.e. can the 
information actually be accessed in ST when necessary?
\end{itemize}

\subsubsection{RE and ST node proportion (P4)}
The proportion of nodes in the RE and ST phases can be used as an indicator for 
the relative effort spent on REST alignment in the respective phases. A REST 
alignment assessment can enable an overview and subsequent optimization by 
removing or changing items detrimental for alignment, as seen next.

\subsubsection{Within-RE (P5a) / Between-Phase links
(P5b) / Within-ST (P5c)}
The amount and quality of Within-RE, Within-ST and Between-Phase links can 
drive improvements that aim to optimize the links as such and their interplay 
throughout the development phases. For example, an ad-hoc connection of 
business with user requirements could be replaced by a more rigorous mechanism 
that explicates relationships between business requirements that are also 
reflected in user requirements. Such a change in RE is however only reasonable 
if user requirements are actually linked to test cases, such that ST can adapt 
its test strategy on the information provided in RE. If the Between-Phase link 
does not exist or is inefficient, an improvement of the RE links would not 
achieve the anticipated benefits. Thus, an SPI initiative or just introducing a 
tool might be beneficial for requirements management, but without a REST 
alignment assessment the benefit or even detrimental nature of a change can not 
be seen.
\begin{itemize}
 \item How does staff turnover affect the quality of requirements and 
derivative artifacts?
 \item In case requirements change, by whom/how/when are these changes 
propagated to linked artifacts?
 \item Does inconsistency of information among artifacts affect the work in: 
RE, ST, the interface between both?
\end{itemize}

\subsubsection{Scope (P6)}
Scope evaluation can lead to the identification of areas in RE or ST where the 
alignment may be improved or even established as the taxonomy makes 
deficiencies in linking information explicit.
Most importantly, it allows to identify gaps that would have not been perceived
as such when looking at them from either the RE or ST perspective alone. For
example, system requirements may be linked to system tests, allowing to
determine test progress. However, if the system requirements are not linked to
business requirements, a statement on the verified portion that provides
business value can not be made.
\begin{itemize}
 \item (Identify artifacts that are created by RE and ST and ask which input is 
used to actually develop them)
 \item How is the consistency between these inputs and the developed artifacts 
maintained over time? What are the advantages/drawbacks of maintaining this 
consistency?
\end{itemize}

\subsection{Case study: REST-bench results}\label{sec:ASResults}
\begin{figure}
  \includegraphics[scale=0.9]{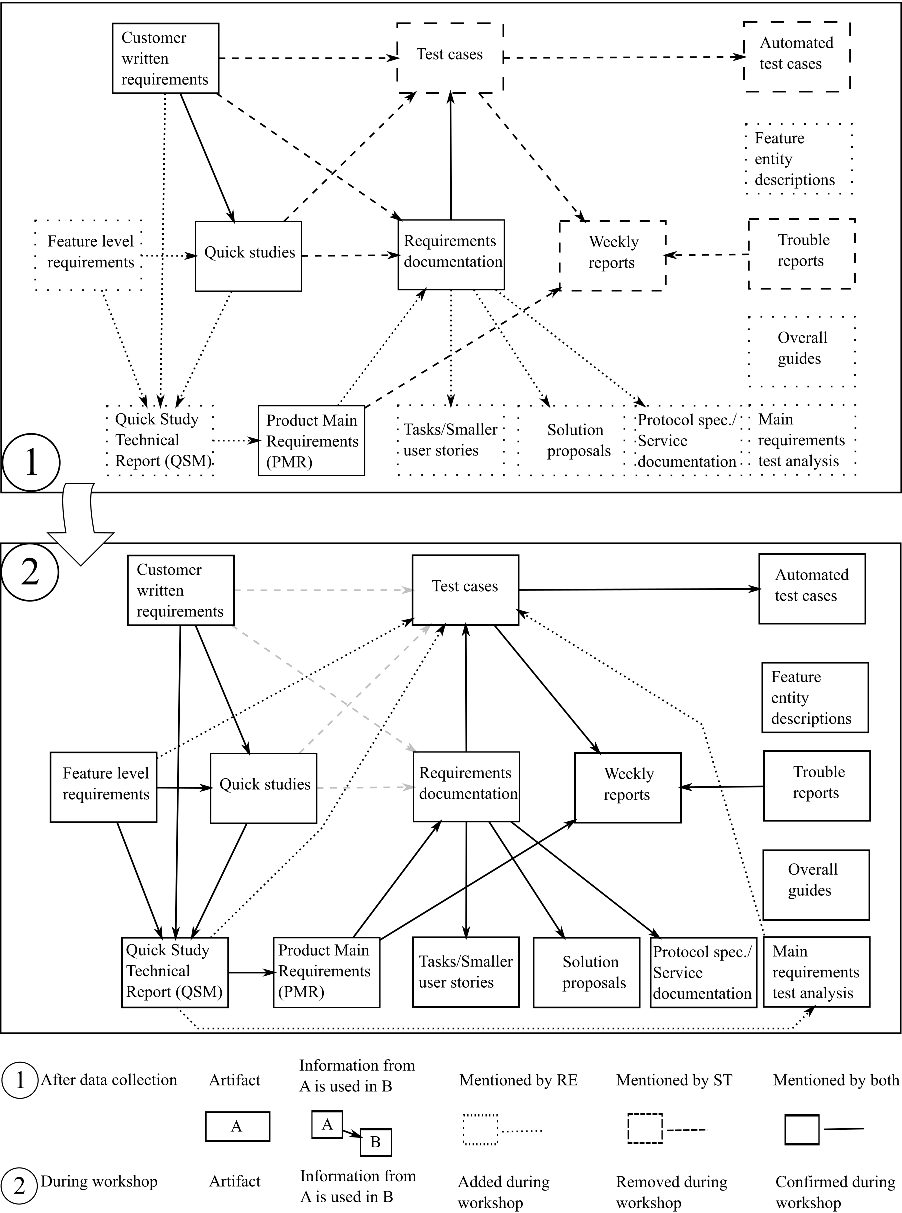}
\caption{REST-bench artifact maps from the Ericsson case study}
\label{fig:AM} 
\end{figure}
The case presented in this paper is based on a research collaboration with
Ericsson~AB, Karlskrona. Ericsson is involved in the development of embedded
applications for the telecommunication domain. The studied project, completed in
autumn 2011, had a duration of appropriately one calendar year. The staff
consisted of 150 engineers which were split up in seven teams. The system
requirements consisted of approximately 350 user stories. The system test cases
amounted to 700, of which 550 were automated. The interviewees were selected
based on their work experience and their collaboration during the project. The
RE representative, a system manager, had 12 years of experience in his current
role whereas the ST representative, a verification engineer, had 14 years of
experience.

Figure~\ref{fig:AM} illustrates the artifact maps that were created after the
interviews (Map~1) and during the workshop (Map~2). To improve readability, the
maps in Figure~\ref{fig:AM} are not annotated with the creator and users of the
artifacts as they were elicited in the interviews. During the workshop these
annotations were however useful to discuss the purpose of certain artifacts
and to identify responsibilities regarding their maintenance during the project
life-cycle. The relationships between artifacts, represented by a directed
arrow, indicate the transfer of information. For example in Map~1, ST uses 
information from ``customer written requirements'', ``quick studies'' and
``requirements documentation'' to create ``Test cases''. Map~1, merged from
the RE and the ST perspective on the project documentation, gave rise to several
observations that were addressed during the workshop:
\begin{longitem}
 \item ST uses more artifacts to create test cases than RE is aware of
 \item ``Feature entity descriptions'' and ``Protocol specifications/Service
documentation'', both created by RE with ST as user, are not used by ST
 \item How are RE artifacts (``Customer written requirements'', ``Feature level
requirements'', ``Quick studies'' and ``Requirements documentation'' kept
consistent when requirements change?
 \item ``Requirements documentation'', consisting of user stories, is not
traced explicitly to the corresponding test cases; a specific role, the
Technical Manager for Test, mediates between test engineers and RE.
\end{longitem}

During the workshop, Map~1 was handed out to the RE and ST representatives.
They reviewed each artifact and link, proposing the changes leading to
Map~2. Notice that most of the changes (removed and added links) refer to the
interface between RE and ST. This indicates that the coordination \emph{within}
RE and ST respectively is well understood, however the interplay \emph{between}
RE and ST is rather opaque when viewed from a single perspective, emphasizing
the value of the process of creating artifact maps. 
The discussion during this process lead also to a knowledge transfer among RE
and ST, clarifying several misconceptions on the use of artifacts by the
different roles in the project:
\begin{longitem}
 \item [Use of artifacts:] RE stated that service documents are rather seldom
updated since they are of little use for ST. ST clarified that they are quite
often used for testing.
\emph{Potential impact:} ST may rely on outdated information in service
documents, leading to failed tests and/or unnecessary trouble reports.
 \item [Lifetime of artifacts:] RE assumed that automated test cases (test cases
for legacy functions) are linked to user stories or requirement statements. ST
clarified that they are mapped to commercial features, which are part of the
product specification and not of the products requirements documentation.
\emph{Potential impact:} The usefulness of the requirements documentation for ST
ends to a certain extent when a manual test case is automated (the link to the
requirements documentation is replaced by a mapping to a commercial feature).
Assuming that requirements are reused in follow-up projects, ST requires
additional effort to decide whether new test cases need to be created or
automated test cases can be reused.
 \item [Information dispersion:] The Technical Manager for Test (TMT) serves as
a link between RE and ST and RE assumed that the TMT initiates testing,
forwarding the necessary information from RE. However, according to ST, testers
actually pull the necessary information from the TMT. 
\emph{Potential impact}: The resources of the TMT are not efficiently used since
he interacts individually with testers. Information may be available to some
testers while others may not have contacted the TMT, leading to inconsistencies
among the testers’ knowledge about the system.
\end{longitem}

\subsection{Identifying improvements using
REST-bench}\label{sec:ASDyadStructureAnalysis}
After the artifact map was reviewed by the workshop participants, leading to
Map~2 in Figure~\ref{fig:AM}, the analyst used the dyad structure properties
(see Section~\ref{sec:REST-benchMethod}) to guide the elaboration of
improvement opportunities on the basis of the artifact map. 

\paragraph*{P1 - Number of nodes} 
Looking at Map~2 in Figure~\ref{fig:AM}, it should be observed that ST uses 
``Feature level requirements'', ``Quick Study Technical Report (QSM)'', ``Main 
requirements test analysis'' and ``Requirements documentation'' as input to 
create test cases. 
The latter one is the main source, whereas the others are used complementary. 
The question arises whether inconsistencies in these documents, caused for 
instance by requirements changes during the design or implementation, affect 
ST. According to RE, changes in user stories (part of the requirements 
documentation) are propagated to the QSM, and the affected development teams 
and ST are invited in presentations where these changes are discussed. In the 
context of this particular project, according to ST, there were inconsistencies 
due to a too early test analysis. RE states that there was a need to redesign 
parts of the solution, leading to consequences which were not dealt with.

\paragraph{P3 - Intermediate nodes}
``Feature entity descriptions'' describe the system functionality on a compound
level. RE states that early creation of this document would help ST since it
shows a better use-case of the system than the documentation written by the
individual development teams (which may be inconsistent with respect to each
other). Feature entity descriptions are written late since they describe how the
system is actually implemented and to be used which is completely known
only quite late in the project. To make the feature entity descriptions useful
to ST, they would need to be maintained and updated during the project as the
implementation stabilizes. RE states that writing the feature entity
descriptions late is a local sub-optimization since other users would benefit of
being able to use them. A factor that may contribute to the difficulty of
maintaining feature entity descriptions is that the responsibility to write them
is given to technical writers (external consultants). 

\paragraph*{P5b - Between-phase links} 
An important interface between RE and ST is the link between ``Requirements
documentation'' and ``Test cases''. There is however no mapping between
requirements and test cases at the Integration Test level. According to ST, this
is due do the difficulty to keep this mapping up-to-date (spreadsheets are
inefficient) or to import the required information into the test management
tool. According to RE this lack of mapping may lead to lower test coverage of
the requirements. 
Looking at Map~2 in Figure~\ref{fig:AM}, the main requirements test analysis
is based on the quick study technical report. The Technical Manager for Test
(TMT) serves as a link between RE and ST, performing the main requirements test
analysis, and defining the scope for the testing effort. In this case, the
between-phase link is implicitly established by a role, leading to the
consequences discussed in Section~\ref{sec:ASResults}, i.e. an inconsistency in
the testers knowledge on the system and test scope.

\subsection{Lessons learned using REST-bench in industry and limitations}
Overall, the assessment can be considered as lightweight in terms of effort for
both the analyst and the organization. The interviewees invested 1.5 hours each
for the interview and 2 hours for the joint workshop. In this first-time 
application of REST-bench, the analysis of the collected interview data,
including the creation of artifact maps and a report summarizing the findings, 
required 5 days of full-time work. We expect however that, with increasing 
experience, the effort for the analysis can be reduced to 2 working days.

The format of the assessment, with separate interviews and a joint workshop,
allowed us to pinpoint disagreements in the perspectives of RE and ST
representatives on project documentation. The test engineer stated that ``we
have a need of improving the coordination between RE and ST. ST is in the end of
the development phase and requirements have been changed/removed and are
sometimes ambiguous. There is always a need to understand what/how/why other
parts of the organization work.'' The requirements engineer agreed that
REST-bench could complement a project post-mortem process, stated however also
that ``it should then include more roles and perhaps also dig into not only
which documents are used but also what [parts of information] in certain
documents are actually used.''

We observed a mutual learning effect among engineers that worked for a year on 
the same project and for over a decade in the same organization. As such, the 
artifacts map and its review during the workshop is a valuable tool to create a 
shared understanding of the coordination mechanisms in a large development 
project. 

The dyad structure properties served as heuristics to identify potential
sources of misalignment between RE and ST. They allow for a focused analysis of
issues that emerge when both RE and ST perspectives are considered. The dyad
structure properties are thereby an useful abstraction of the detail in the
REST taxonomy, enabling a more effective interaction and communication with
industry.

In the assessment we used artifacts (their creation and use) as a proxy to
elicit and understand the alignment between RE and ST. Artifacts are tangible
and relatively easy to describe in terms of their purpose and content. Hence
they were a natural choice to structure the elicitation and analysis. The
assessment can be however extended to achieve a more fine-grained picture of the
state of REST alignment. For example, informal communication channels and
information sources, such as e-mail, internal wikis, instant messaging and
telephone calls, or meetings could be included in the analysis. Such a detailed
assessment would however require to focus the analysis on a limited set of
activities where RE and ST interact.

\section{Conclusion and future work}\label{sec:Conclusion}
Taxonomies are means to structure our knowledge and to discover new
relationships between concepts, advancing the understanding of observed
phenomena. The taxonomy presented in this paper aims at characterizing methods
for requirements engineering (RE) and software test (ST) alignment. Although
both RE and ST are mature research areas, their interplay has been less
explored. Investigating these relationships, structuring and communicating them
are the major contributions as summarized below:
\begin{longitem}
 \item We have investigated the RE and ST alignment phenomenon by applying a
bottom-up and iterative approach to construct a taxonomy. The principles of
this approach might also be useful to construct taxonomies in other research
areas.
 \item The structuring of concepts that belong to different areas is a
challenging task. The information dyad is an abstraction that supports the
reasoning on RE and ST alignment methods. Although we expect that the taxonomy
can be refined and extended, we claim that the concept of an information dyad is
a valid construct to characterize RE and ST alignment in particular, but we
also foresee it as being valid to characterize alignment in other domains of SE
as well. Essentially, information can never be adapted or aligned to each
other without some type of link between them. The dyad makes this into
a first class concept and thus allows to clarify and compare links as
well as the information being linked much more explicitly and formally.
 \item Clear definitions support communication. We have defined RE and
ST alignment both as a state and as an activity, since the meaning may differ
depending on the context. In the context of alignment-as-activity, we developed 
the REST taxonomy, identifying and describing dyad structure properties that 
allow one to reason upon the phenomenon of alignment. In the context 
of alignment-as-state, we developed REST-bench, an alignment assessment 
framework powered by the REST taxonomy.
\end{longitem}

The application of the taxonomy on 13 REST alignment methods allowed us to
reason on the overall topology of the methods. We have observed a median of
three information dyads and a tendency to have more nodes in earlier development
phases, i.e. in RE, than in the ST phase. Assuming that the complexity of a 
method, and therefore the effort of applying it, increases with the number of 
nodes, this indicates that most of the classified alignment methods require a 
relatively higher effort in the start-up phase (RE) than in follow-up phases 
(Analysis, Implementation, ST). Interestingly, the two most complex alignment 
methods (Case E and C), according to their dyad structure properties, are geared 
towards a focus in ST. This is rather surprising as one would intuitively drive 
the alignment effort from requirements engineering activities (this intuition is 
confirmed by the majority of the other methods we classified). This could 
indicate that these two methods work better in a context where the RE activities 
relevant for alignment are already well understood and established. 

On the opposite end of the spectrum, we observed four alignment methods (Case A, 
K, M and I) that were classified with a similar, low complexity, differing 
however in their focus and scope. However, only Case K has a scope that reaches 
from early requirements engineering to late testing. This indicates that the 
complexity of alignment methods correlates with their scope (although we 
observed also an exception with Case F), which is not surprising. 

Based on the manual search in 635 publications we identified only five REST
alignment methods. One explanation for this low number may be the strict
inclusion criteria, i.e. that the publication fulfills the relevance criteria
(scope, comprehensiveness, rigor) stated in Section~\ref{sec:ProcessRelevance}. 
The scope criterion turned out to be the most selective as it excluded a wide 
range of publications that did not target \emph{both} the RE and ST areas (see 
Section~\ref{sec:Iteration5}). Even though the analysis of the studied methods 
revealed that the \emph{connection} mechanism, which enables traceability, was 
found most frequently, the overall low number of identified REST alignment 
methods signals for more research and solution proposals that aim to bridge the 
gap between RE and ST.

We developed an assessment framework with the REST taxonomy as engine, called
REST-bench, and applied it in an industrial assessment at Ericsson AB. The
concepts of the REST taxonomy, integrated in REST-bench, turned
out to be useful in transferring knowledge between RE and ST roles and in
clarifying misunderstandings between them. By representing the coordination
between RE and ST in an artifact map, we used the same heuristics of the dyad
structure  analysis performed earlier on the alignment methods, leading
to the identification of bottlenecks (i.e. synchronization of too many
artifacts) and sub-optimizations (i.e. late creation of artifacts) in the
interaction between RE and ST. Since REST-bench is very lightweight,
it could be integrated into post-mortem procedures that many organizations
typically perform after the closure of a development project.               
                          
The REST taxonomy provides a novel view on the aligning requirements
engineering and software test, based on a rigorous classification scheme and
method. It enables the characterization of alignment methods and the
assessment of alignment in a development organization. We are continuing our
work on utilizing REST-bench as an alignment assessment aid, since the
underlying taxonomy and the developed dyad structure properties have a great
potential of identifying, characterizing and probing strengths and weaknesses in
the alignment between RE and ST in industry, in particular as a sanity check for
process improvement initiatives.

\appendix
\section*{APPENDIX}
\setcounter{section}{1}

\subsection{Taxonomy application on alignment methods -
Results}\label{sec:AppendixTaxAppMethodResults}

\begin{table}[t]\footnotesize
 \caption{Results of the REST taxonomy application on four in-depth cases}
 \begin{centering}
 \subfloat[Case A]{\label{subtab:case_A}
  \begin{minipage}[c]{0.28\textwidth}
   \includegraphics[scale=1]{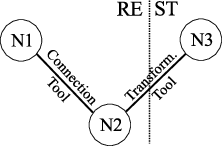}
  \end{minipage}
  \begin{minipage}[c]{0.72\textwidth}
   \centering
   \begin{tabular}{lp{3.5cm}p{3.5cm}p{1.8cm}}\toprule
    ID & Node name & Information & Owner\\ \midrule
    N1 & Requirement specification & Nat. lang. specifications & $\emptyset$\\
    N2 & Requirements analysis & Requirements clusters & $\emptyset$\\ 
    N3 & Test specification & Testplan / Abstract test cases & Test manager\\
   \midrule
   \multicolumn{4}{l}{Dyads (2): N1-N2, N2-N3}\\
   \bottomrule
   \end{tabular}
  \end{minipage}
 }
 \end{centering}
 \\
 \begin{centering}
 \subfloat[Case B]{\label{subtab:case_B}
  \begin{minipage}[c]{0.32\textwidth}
   \includegraphics[scale=1]{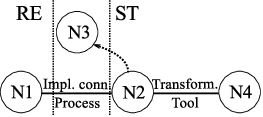}
  \end{minipage}
  \begin{minipage}[c]{0.68\textwidth}
   \centering
   \begin{tabular}{lp{3.5cm}p{4cm}p{0.7cm}}\toprule
    ID & Node name & Information & Owner\\ \midrule
    N1 & Requirement specification & System functional requirements &
$\emptyset$\\
    N2 & Test modeling & Abstract test cases & $\emptyset$\\ 
    N3 & Implementation & System specific configuration & $\emptyset$\\ 
    N4 & Test implementation & Executable test cases & $\emptyset$\\
   \midrule
   \multicolumn{4}{l}{Dyads (2): N1-N2, N2-N4}\\
   \bottomrule
   \end{tabular}
  \end{minipage}
 }
 \end{centering}
 \\
 \begin{centering}
 \subfloat[Case C]{\label{subtab:case_C}
  \begin{minipage}[c]{0.25\textwidth}
   \includegraphics[scale=1]{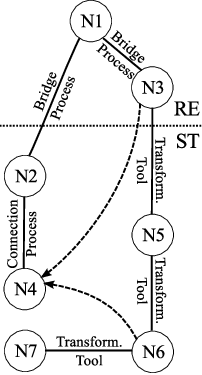}
  \end{minipage}
  \begin{minipage}[c]{0.75\textwidth}
   \centering
   \begin{tabular}{lp{4cm}p{3.5cm}l}\toprule
    ID & Node name & Information & Owner\\ \midrule
    N1 & Requirement specification & Nat. lang. specifications & $\emptyset$\\
    N2 & Scenario definition & Factors and Levels & Test designer\\ 
    N3 & Requirement analysis & Dictionary & Test designer\\ 
    N4 & Scenario mapping & SRS to scenario mapping & Test designer \\
    N5 & Test modeling & Statechart model & Test designer \\
    N6 & Test specification & Abstract test cases & Test designer \\
    N7 & Test implementation & Executable test cases & Test designer \\
   \midrule
   \multicolumn{4}{l}{Dyads (6): N1-N2, N1-N3, N2-N4, N3-N5, N5-N6, N6-N7}\\
   \bottomrule
   \end{tabular}
  \end{minipage}
 }
 \end{centering}
 \\
 \begin{centering}
 \subfloat[Case D]{\label{subtab:case_D}
  \begin{minipage}[c]{0.35\textwidth}
   \includegraphics[scale=1]{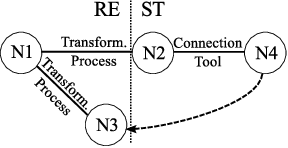}
  \end{minipage}
  \begin{minipage}[c]{0.65\textwidth}
   \centering
   \begin{tabular}{lp{3.1cm}p{3.1cm}l}\toprule
    ID & Node name & Information & Owner\\ \midrule
    N1 & Requirement specification & Use cases, domain models & Bus.
analyst\\
    N2 & Acceptance test design & High level acceptance tests & Bus.
analyst\\ 
    N3 & Requirement analysis & Robustness diagrams & Bus.
analyst\\ 
    N4 & Acc. test implementation & Executable acceptance tests & Bus. analyst
\\
   \midrule
   \multicolumn{4}{l}{Dyads (3): N1-N2, N1-N3, N2-N4}\\
   \bottomrule
   \end{tabular}
  \end{minipage}
 }
 \end{centering}
 \label{tab:TaxApplication}
\end{table}

\begin{table}[t]
\tbl{Context of the four in-depth cases\label{tab:CaseContext}}{%
 \begin{tabular}{lp{11cm}}
  \toprule
  \multicolumn{2}{c}{Case A}\\
  \toprule
  Aspect & Description\\ 
  \midrule
  Method setting &  319 requirements, functional and non-functional
requirements, requirements-based acceptance testing / test-planning, bespoke RE,
natural language requirements \\ 
  Focus & 4) Main purpose was to improve / affect alignment\\ 
  Motivation & Inconsistencies, redundancies and dependencies in requirements
documents, representing different viewpoints of a system, lead to erroneous
tests, increased test effort and complex test plans \\
  Assumptions & Natural language requirements specified in template-form\\ 
  Quality targets & Improve test plans and make the test process more efficient
\\
  Validation & Clustered requirements and generated test-plans are reviewed
and, if necessary, refined \\
  \bottomrule 
\\
  \multicolumn{2}{c}{Case B}\\
  \toprule
  Aspect & Description\\ 
  \midrule
  Method setting & 1500 abstract and 200,000 executable tests, implementation
and verification under safety norms, functional requirements, system testing
(instantiation of abstract test cases based on concrete system configurations)
\\ 
  Focus & 4) Main purpose was to improve / affect alignment\\ 
  Motivation & Manual instantiation of configuration-specific test cases is a
time-consuming activity \\
  Assumptions & Requires a certain reference software architecture; control
system can be abstracted as a FSM\\
  Quality targets & Improve configuration coverage in system tests \\
  Outcome & Generation of executable test cases is by orders of magnitude
more efficient than creating them manually \\ 
  \bottomrule 
\\
  \multicolumn{2}{c}{Case C}\\
  \toprule
  Aspect & Description\\ 
  \midrule
  Method setting & 175 scenarios derived from requirements, SDLC includes
independent V\&V, functional requirements, model-based system and acceptance
testing, bespoke requirements, natural language requirements\\ 
  Focus & 5) Intended, main as well sole purpose\\ 
  Motivation &  Natural language used the most in specifying software
requirements and deriving scenarios for system and acceptance tests is
challenging and time-consuming\\
  Outcome & Quality of derived executable test cases is comparable to expert's\\
  \bottomrule 
\\
  \multicolumn{2}{c}{Case D}\\
  \toprule
  Aspect & Description\\ 
  \midrule
  Method setting &  small-scale example application, functional requirements,
acceptance testing, natural language requirements and models \\ 
  Focus & 5) Intended, main as well sole purpose\\ 
  Motivation & Lack of process that allows analysts to develop acceptance
tests from use case models without requiring additional design artifacts\\
  Quality targets & Development of comprehensive and effective acceptance
tests\\
  Validation & With robustness diagrams, consistency of use case and domain
models can be checked informally \\
  \bottomrule
\end{tabular}}
\begin{tabnote}
 \Note{Note:}{Aspects which could not be identified in the paper are not shown
in the table.}
\end{tabnote}
\end{table}

The application of the taxonomy on 4 cases is summarized in
Tables~\ref{tab:TaxApplication} and~\ref{tab:CaseContext}.
Table~\ref{tab:TaxApplication} illustrates the characterization of the
identified information dyads. The empty-set symbol ($\emptyset$) indicates that
we could not identify the owner of information in the respective node. The
graphs on the left in Table~\ref{tab:TaxApplication} illustrate the dyad
structures, showing also the mechanism and medium characterizing the link.
Table~\ref{tab:CaseContext} summarizes the respective method context. In
the following paragraphs we describe each case, motivating the identified
dyad structures. 
\paragraph*{Case A} \citeN{guldali_torc:_2011} present a method and tool
support for deriving optimized acceptance test plans from requirements
specifications. 

We have identified three nodes as illustrated in Table~\ref{subtab:case_A}.
The information in N1 is multi-viewpoint requirements specifications, capturing
different aspects of the software system and its environment. Since these
aspects contain overlaps, the requirements specifications may be redundant.
Leveraging on linguistic analysis, similar requirements are clustered (N2) such
that a reduced set of test steps and asserts can be derived (N3).
The mechanism in dyad N1-N2 is a \emph{connection} since notation and meaning of
the information are not changed (individual requirements are mapped into
clusters). In dyad N2-N3 we observe however a \emph{transformation} mechanism
since the relationship between requirements is preserved in the derived test
plan, expressed in the order of the test steps. 

\paragraph*{Case B} \citeN{flammini_automatic_2009} propose a method to automate
the verification of computer-based control systems on different configurations.

As shown in Table~\ref{subtab:case_B}, we have identified four nodes. The
mechanism in dyads N1-N2 is an \emph{implicit connection} since the description
of the process does not reveal how abstract test cases are derived from the
functional requirements (the focus of the study is on the instantiation of
abstract test cases). For the derivation of executable test cases (N4), the
method depends on abstract test cases (N2), and uses the configuration data
specific for the system under test (N3). The ``use'' relationship is shown by
the dashed link between N2 and N3.
The mechanism in dyad N2-N4 is a \emph{transformation} since the notation of
the non-executable test model (abstract test cases) is translated into
concrete test cases, preserving the relationships between the abstract test
cases by leveraging the information provided in the system specific
configuration (N3).

\paragraph*{Case C} \citeN{de_santiago_junior_generating_2012} describe a
methodology to generate scenario-based test cases from natural language
requirements.

We have identified seven nodes as illustrated in Table~\ref{subtab:case_C}.
The information in N2, factors and levels upon which scenarios are generated,
are derived from the software requirements specification (N1). The link
mechanism is a \emph{bridge} since in the process, the notation of the
information in N2 changes, but the meaning (i.e. the decision whether a certain
combination of factors and levels is relevant) has to be maintained by the
test designer. The analogous argument applies for the dyad N1-N3, in which the
test designer needs to apply his domain knowledge to establish the dictionary
(N3).
In dyad N2-N4, the test designer establishes a mapping between the generated
test scenarios and the related requirements. This information is used (hence
the dashed connector) in N3 and N6 to generate statechart models and executable
test cases. Since the link in dyad N2-N4 establishes a logical link between
requirements and scenarios, we define the mechanism as a \emph{connection}. 
The dyads N3-N5, N5-N6 and N6-N7 feature a \emph{transformation} mechanism since
relationships between requirements (stemming from the scenario mapping in N4)
are preserved in the statechart model (N5), in the abstract test cases (N6) and
in the executable test cases (N7).

\paragraph*{Case D} \citeN{el-attar_developing_2010} propose a method to derive
executable acceptance tests from use case models, domain models and robustness
diagrams.

We have identified four nodes as shown in Table~\ref{subtab:case_D}. In dyad
N1-N2, the business analyst derives from use case and domain models high level
acceptance tests (HLAT) which can be used to evaluate the system manually. We
observe a \emph{transformation} mechanism since the procedure for
maintaining the relationships between use cases in the corresponding HLATs is
well defined. Similarly, robustness diagrams (N3) are constructed to
ensure consistency between use cases and domain models. The link in dyad N2-N4
is established by a \emph{connection} mechanism, mapping HLATS with executable
test cases and using robustness diagrams (dashed connector N4-N3).

\bibliographystyle{acmsmall}
\bibliography{p3}

\end{document}